\documentclass{jfm}
\usepackage[utf8]{inputenc}
\usepackage[breaklinks]{hyperref}
\usepackage{graphicx}

\usepackage{natbib}
\usepackage{outlines}
\usepackage{hyperref}

\usepackage{indentfirst}
\usepackage{amsmath}
\usepackage{xfrac}
\usepackage{subfigure}

\usepackage[usenames,dvipsnames]{pstricks}
\usepackage{epsfig}
\usepackage{pst-grad} 
\usepackage{pst-plot} 

\ifCUPmtlplainloaded \else
  \checkfont{eurm10}
  \iffontfound
    \IfFileExists{upmath.sty}
      {\typeout{^^JFound AMS Euler Roman fonts on the system,
                   using the 'upmath' package.^^J}%
       \usepackage{upmath}}
      {\typeout{^^JFound AMS Euler Roman fonts on the system, but you
                   dont seem to have the}%
       \typeout{'upmath' package installed. JFM.cls can take advantage
                 of these fonts,^^Jif you use 'upmath' package.^^J}%
      }
  \else
  \fi
\fi

\ifCUPmtlplainloaded \else
  \checkfont{msam10}
  \iffontfound
    \IfFileExists{amssymb.sty}
      {\typeout{^^JFound AMS Symbol fonts on the system, using the
                'amssymb' package.^^J}%
       \usepackage{amssymb}%

      }{}
  \fi
\fi

\ifCUPmtlplainloaded \else
  \IfFileExists{amsbsy.sty}
    {\typeout{^^JFound the 'amsbsy' package on the system, using it.^^J}%
     \usepackage{amsbsy}}
    {}
\fi

\newsavebox{\astrutbox}
\sbox{\astrutbox}{\rule[-5pt]{0pt}{20pt}}

\usepackage{ulem}

\def\ADD#1{{\textcolor{black}{#1}}}    
\usepackage{graphicx}
\usepackage{color}

\title[Internal wave attractors:  \ADD{different scenarios of instability}]
{Internal wave attractors: \ADD{different scenarios of instability}}
\author[%
  C. Brouzet,  E. Ermanyuk, S. Joubaud, G. Pillet,  T. Dauxois]
  {%
    C. Brouzet\textsuperscript{1},  E. Ermanyuk\textsuperscript{1,2}, S. Joubaud\textsuperscript{1}, G. Pillet\textsuperscript{1} and T.~Dauxois\textsuperscript{1}
}

\affiliation{
  \textsuperscript{1}
 Univ Lyon, ENS de Lyon, Univ Claude Bernard, CNRS, Laboratoire de Physique, F-69342 Lyon, France\\
  [\affilskip]
  \textsuperscript{2}
  Lavrentyev Institute of Hydrodynamics, av. Lavrentyev 15, Novosibirsk 630090,
  Russia \\
  }
\pubyear{???} \volume{???} \pagerange{???}
\date{\today}
\begin{document}

\maketitle

\begin{abstract}

This paper presents an experimental study \ADD{of different instability scenarios in} a paralle\-logram-shaped 
 internal wave attractor in a trapezoidal domain filled with a uniformly stratified fluid.
Energy is injected into the system via the oscillatory motion of a vertical wall of the trapezoidal domain. 
Whole-field velocity measurements are performed with the conventional PIV technique.  

\ADD{In the linear regime, the total kinetic energy
of the fluid system is used to quantify the strength of attractors as a function of coordinates in the parameter space
defining their zone of existence, the so-called Arnold tongue.} \ADD{In the nonlinear regime,} the choice of 
the operational point in the Arnold tongue is shown to have a significant impact on the scenario of the 
onset of triadic instability, most notably on the
 \ADD{influence} of confinement on secondary waves.
The onset of triadic resonance instability may occur as a spatially localized event similar to~\cite{SED2013} in the case of strong focusing  
or in form of growing normal modes as in~\cite{McEwan1971} for the limiting case of rectangular domain. In the present paper, we describe 
also a new intermediate scenario for the case of weak focusing.

We explore the long-term behaviour of cascades of
triadic instabilities in wave attractors and show a persistent trend toward formation of standing-wave patterns
corresponding to some discrete peaks of the frequency spectrum. At sufficiently high level of energy injection 
the system exhibits a "mixing box" regime which has certain qualitatively universal properties regardless to the choice
of the operating point in the Arnold tongue. In particular, for this regime, we observe a 
 statistics
of events with high horizontal vorticity, which serve as kinematic indicators of mixing.
\end{abstract}

\section{Introduction.}

Internal waves are ubiquitous in large geophysical systems such as oceans, seas and lakes. All these systems are geometrically confined. The importance of confinement depends on the ratio between the scale of the internal wave motion and the size of the domain, and on the rate of energy dissipation. Typical limiting cases include purely standing waves in a fully confined reservoir (e.g. internal seiches in lakes~\citep{Hutteretal2011}), and purely \ADD{propagating} waves in laterally unbounded domains (e.g. baroclinic tides emitted by an isolated bottom topography~\citep{GarrettKunze2007}). Experimental and geophysical reality often lies inconveniently between these two extremities. 

Dynamics of internal waves in confined domains filled with continuously stratified fluid is significantly enriched by the specific form of the dispersion relation. Uniformly  stratified fluid (with constant buoyancy frequency $N$), a conventional simplification for a continuous stratification, supports internal waves propagating 
in form of oblique beams {with the phase (group) velocity vector} tilted at an angle $\theta$ to the vertical {(horizontal)}
 defined by the dispersion relation $\sin \theta= \pm\, \omega/N=\pm\,\Omega$, where $\omega$  is the forcing frequency, and $\Omega$ is its nondimensional counterpart normalized by  $N=[(-g/\bar{\rho})({\rm d}\rho/{\rm d}z)]^{1/2}$, with $g$ the gravity acceleration, $\rho (z)$ the density distribution along the vertical coordinate~$z$ and~$\bar{\rho}$ a reference value~\citep{MowbrayRarity1967,Turner1973}. 

This anisotropic dispersion relation represents a strong geometric restriction since it requires preservation of the angle  $\theta$ to the {horizontal} upon reflection of the wave beam at a rigid wall. For vertical or horizontal walls, reflection is similar to the classic case of specular reflection considered in geometrical optics. Accordingly, a typical internal wave regime in rectangular domains is represented by standing waves (normal modes) as described in~\cite{Turner1973}. These waves can be obtained as a sum of four identical monochromatic (in time and space) \ADD{propagating} waves corresponding to four possible choices of~$\theta$ in the domain with depth (resp. length) equal to integer number of half-wavelength in vertical (resp. horizontal) direction. Considering the limit of infinitely thin wave beams as geometric rays, one obtains the web of rays mapped exactly on themselves upon reflection (global resonance). 

If the confined system has a sloping wall, focusing/defocusing occurs~\citep{DauxoisYoung1999}, the width of the wave beam decreases/increases upon reflection. Assuming a certain particular geometry in a two-dimensional problem, one can obtain a full classification of the observed wave regimes as it is done for parabolic and trapezoidal domains \citep{MaasLam1995,MBSL1997}. Careful investigation shows that focusing prevails: internal wave rays converge to closed trajectories called internal wave attractors. A convenient classification of the observed regimes can be done~\citep{MBSL1997} in terms of Lyapunov exponents that characterize quantitatively the rate of convergence of the ray trajectories toward limiting cycles. The Lyapunov exponents can be plotted as a function of two parameters, where one parameter $d$ controls the shape of the domain while another parameter
$\tau$ controls the {geometric ray pattern for a given $d$} (see figure~\ref{nouveaudiagram}). By using the idea of affine similitude, the control {over the ray pattern} can be achieved by the change of $\Omega$ or by rescaling the depth of fluid.  {The map of the Lyapunov exponents as a function of these two parameters shows clearly defined regions of strong convergence, the so called  \ADD{
Arnold tongues \ADD{\citep{MandersMaas2003,Maas2005}} (appearing e.g. in the study of the Mathieu equation) corresponding to wave attractors of different complexity. The Arnold tongues are separated by regions of low convergence where the Lyapunov exponents exhibit a fan-shaped pattern of alternating dark- and light-grey regions, where dark regions correspond to almost zero Lyapunov exponents. It is important that the map contains embedded lines corresponding to global resonances (seiche modes). At these lines, the Lyapunov exponents are identically zero so that the ray pattern is mapped exactly on itself. It should be stressed that the pattern of dark-grey regions is much more dense that the pattern of lines corresponding to global resonance.}

\begin{figure}
   \begin{centering}
  \includegraphics[width=\textwidth]{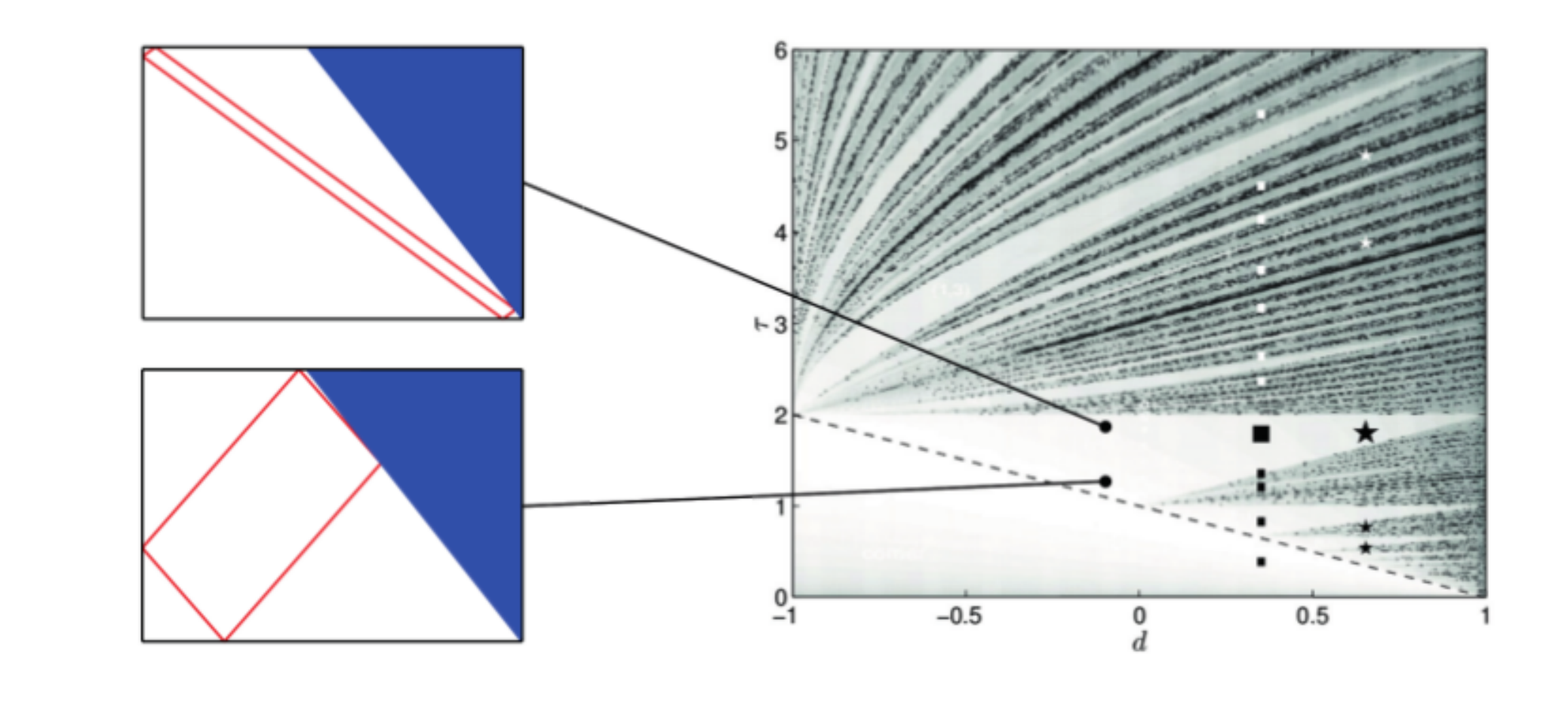}
  \caption{Right panel presents the ($d$,$\tau$) diagram obtained by \cite{MBSL1997}. 
  Left panels present two examples of the corresponding (1,1) attractor.
  The  different symbols 
 in the right panel correspond to the different peak frequencies for the 2 experiments ($\blacksquare$ and $\bigstar$ in Table~2 with $a=5$~mm). The primary wave frequency is plotted with a filled and larger symbol.
 The subharmonics (respectively superharmonics) are represented by white (respectively black) symbols. 
    \label{nouveaudiagram}}
   \end{centering}
\end{figure}

The concept of wave attractor emerging as a result of ray tracing assumes that the attractor represents a 
\ADD{propagating} wave. In a viscous fluid, the width of the attractor beams is set by the
 competition between geometric focusing and viscous broadening~\citep{Ogilvie2005,HBDM2008,GSP2008} 
 what once again assumes a \ADD{propagating} wave: one can even put forth an analogy 
 with a wave beam emitted by an oscillating object as discussed in~\cite{GSP2008} and \cite{JouveOgilvie2014}. 
This concept works well if the width-to-length ratio for all beams of the attractor is small. This "narrow-beam"
 approximation is violated for geometrically degenerated systems: for example, when the shape of the fluid 
domain does not support focusing (a rectangular domain) or when the attractor becomes very elongated and
ultimately collapses onto a line (see figure~\ref{nouveaudiagram}). 

In the present paper, we consider experimentally the effect of geometric degeneracy on the structure of the Arnold tongue for a generic case of $(1,1)$ attractor in trapezoidal geometry. $(1,1)$ attractor represents a parallelogram with one reflecting point at each side of the trapezoid~\citep{MBSL1997}. This attractor emerges when the forcing frequency falls into the range whose upper and lower limits are defined by the slopes of the diagonals of the trapezoid. The diagonals represent the degenerated forms of $(1,1)$ attractor. In addition, the trapezoidal geometry does not support focusing when it degenerates into a rectangle. 
\ADD{To investigate the strength of attractors as a function of their location in the Arnold tongue}, we introduce the total kinetic energy of the confined system as a variable, which allows to quantify the "susceptibility" of the system to forcing and to discriminate between nearly standing and nearly \ADD{propagating} waves.

Further, we explore the scenario of the wave instability via triadic resonance as a function of the operational 
point of the experimental system at the Arnold tongue: as introduced in~\cite{Brouzetetal2016},
we will use the acronym TRI for the abbreviation of Triadic Resonance Instability. 

Two scenarios of TRI in a confined fluid domain have been described in literature so far. \begin{itemize}
\item[i)] TRI in a rectangular basin has been studied theoretically and experimentally 
in~\cite{McEwan1971}. 
The effect of confinement in this case is fully present since all waves in the resonant triad are assumed to be standing. 
\item[ii)] 
TRI in a trapezoidal domain has been studied experimentally  for $(1, 1)$ attractors in~\cite{SED2013} and numerically in~\cite{Brouzetetal2016}.  They describe the onset of the resonance as a local event in the most energetic branch of the attractor similar to~\cite{KoudellaStaquet2006} and \cite{BDJO2013}, with the effect of finite width of the wave beam involved~\citep{KarimiAkylas2014,BSDBOJ2014}. The primary and secondary wave vectors have been measured and no clear effect of the confinement has been detected. Note that all waves involved into the measured triad are \ADD{propagating}. Thus the confinement in this scenario appears only indirectly as a factor defining the overall shape of the attractor and the width of the primary wave beam in accordance with the mechanism revealed by~\cite{Ogilvie2005}, \cite{HBDM2008} and \cite{GSP2008}.
\end{itemize}

In this paper, we show that the choice of the operational point in the Arnold tongue allows to observe the scenario of instability that is intermediate between the two above-mentioned cases. Further, we analyse the properties of the wave regimes which develop as result of triadic instability by considering their time-frequency diagrams, Hilbert transforms and probability density functions (PDFs) of horizontal vorticity.

The present paper is organized as follows. In Section 2, we describe the experimental set-up. In Section 3, we describe the structure of the Arnold tongue of a stable $(1,1)$ attractor. Section 4 presents three scenarios of the onset of TRI as a function of the choice of the operating point of the experimental system in the Arnold tongue. We describe in detail a new scenario which is intermediate between the cases considered in \cite{SED2013} and \cite{McEwan1971}. This new scenario shows the emergence of standing secondary waves due to weak TRI in a weakly-focused attractor. We analyse then in Section 5 the long-term behaviour of a cascade of TRI in a {well-}focused attractor. We demonstrate that some secondary waves in the rich wave pattern emerging as a result of the cascade correspond to quasi-global resonances (standing waves). The results presented in Sections 4 and 5 show a strong general trend toward emergence of standing-wave components as a result of a simple one-stage triadic resonance or more sophisticated multi-stage  TRI cascades. Finally, in Section 6, we consider the statistical properties of the internal wave regimes obtained as a result of well-developed cascades of triadic interactions. We show that at sufficiently large forcing the system reaches the regime of an ``internal wave mixing box". The statistics of extreme events in the system and the resulting mixing efficiency can be efficiently controlled by the amount of the input energy and by the choice of the operating point in the Arnold tongue. The main findings of this study are summarized in Section 7.

\section{Experimental set-up}

The experimental set-up used in the present work and sketched in figure \ref{setup} is similar to the one described in \cite{SED2013} and~\cite{Brouzetetal2016}. Experiments are conducted in the rectangular test tank of size $800\times 170 \times 425 ~$mm$^{3}$ filled with uniformly stratified fluid using the conventional double-bucket technique. Salt is used as a stratifying agent. The density profile is measured prior and after experiments by a conductivity probe attached to a vertical traverse mechanism. The value of the buoyancy frequency $N$ is evaluated from the measured density profile.
The trapezoidal fluid domain of length~$L$ (measured along the bottom) and depth $H$ is delimited by a sliding sloping wall, inclined at the angle $\alpha$. The wall is slowly inserted into the fluid after the end of the filling procedure. \ADD{The global Archimedes number representing the ratio of the Reynolds and Froude numbers based on the fluid depth is kept fixed throughout all the experiments so that $Ar=H^{2} N/\nu=85000$, where $\nu$ is kinematic viscosity. Thus, at the global scale of the experimental setup the effect of viscosity is reasonably weak, allowing an energy cascade toward smaller dissipative scales.}

The input forcing is introduced into the system by an internal wave generator described in \cite{Gostiaux:EF:07}, \cite{MMMGPD2010} and \cite{JMOD2012}. The time-dependent vertical profile of the generator is prescribed in the form
$$
\zeta(z,t)=a\sin(\omega_{0} t) \cos(\pi z/H),
$$
where $a$ and $\omega_{0}$ are the amplitude and frequency of oscillations, respectively. In a horizontally semi-infinite domain, the motion of the generator would generate the first vertical mode of internal waves. The profile is reproduced in discrete form by the horizontal motion of a stack of $47$ plates.

The whole-field velocity measurements are performed via the standard PIV technique. The fluid is seeded with light-reflecting {hollow glass spheres}
of size $8~\mu$m and density $1100$~kg/m$^{3}$. The sedimentation velocity of particles is found to be very low, with negligible effect on results of velocity measurements. The longitudinal \ADD{$(x,y=0,z)$} mid-plane of the test section is illuminated by a vertical laser sheet coming through the side of the tank. The velocity field is calculated with the help of the cross-correlation technique (Fincham and Delerce 2000) with a typical resolution of 1 velocity vector per area of about $3\times 3$~mm$^2$.  The mesh of measurements is found to be sufficient to resolve the small-scale details of the wave field.

\begin{figure}
   \begin{centering}
   \includegraphics[width=0.5\textwidth]{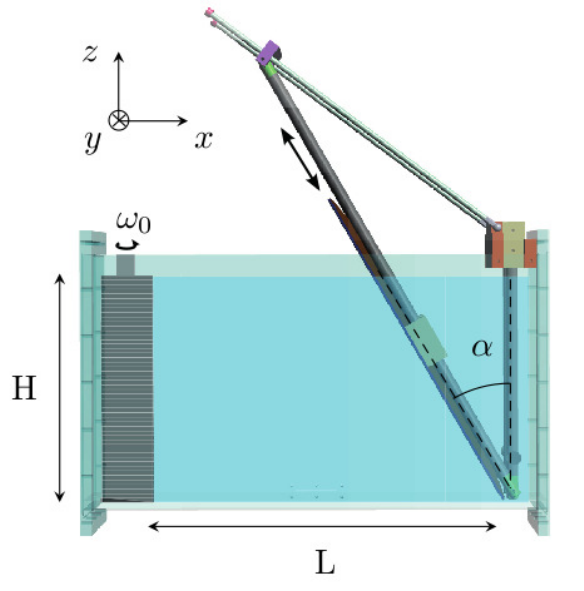}
  \caption{Sketch of the experimental set-up showing the wave generator and the sloping wall (inclined at an angle~$\alpha$ with respect to the vertical)
inside
the immobile tank
  of size $800 \times170 \times 425$~mm$^3$. The working
bottom length of the section, the depth and the sloping angle are given within the text.
   \label{setup}}
\end{centering}
\end{figure}

A computer-controlled video  AVT (Allied Vision Technologies) Stingray {or Pike} camera  with CCD matrix of $2452 \times2054$ pixels is used for video recording of the fluid motion. The camera is located at a distance of $2850$~mm from the test tank. 
For the PIV measurements in short-term experiments, the camera operates at the constant frame rate of 4 Hz. After the PIV treatment, $4$ velocity fields per second are obtained, yielding typically around $40$ fields per period of forcing. 
In long-term experiments, the camera operates in burst mode taking two images separated by $0.125$~s and waiting $0.375$~s before the subsequent burst. The PIV treatment in this case yields $2$ velocity fields per second. This sampling rate is found to provide a sufficient resolution of the significant frequency components of the signal. 

\section{Arnold tongue of a stable $(1,1)$ attractor} { 

To \ADD{quantify the strength of $(1,1)$ attractor}, we consider the total dimensionless kinetic energy of the fluid confined in the trapezoidal domain. Experimentally, this quantity is measured in the vertical longitudinal mid-plane and defined as follows
\begin{equation}
 K=\frac{\displaystyle \int_S\mbox{d} x\mbox{d} z\, \frac{1}{2}(v_x^2+v_z^2)}{\displaystyle\frac{1}{2}(a \omega_0)^2 S},
\end{equation}
where $v_x$ and $v_z$ are the horizontal and vertical velocity components and  $S$ is the area of the trapezoidal domain.  \cite{Brouzetetal2016} show that the internal wave pattern in the experimental set-up is reasonably two-dimensional. Thus, the measurements performed in the vertical mid-plane are representative for the whole volume except thin near-wall boundary layers.

Figure~\ref{kinetic_energy_vs_time} presents a  typical experimental time-history of $K$. It can be seen  that after a transient, which has a typical duration of about $25$~periods, the process reaches saturation and kinetic energy oscillates between certain well-defined minimum  and maximum values. We denote its corresponding time-averaged value as $\left\langle K\right\rangle$, performed in the saturated regime. The quantity $\left\langle K\right\rangle$ can be interpreted as a susceptibility, \ADD{i.e. a dynamic response} of the \ADD{experimental} system to the prescribed forcing of unit amplitude $a$. In addition, one can introduce $R=K_{min}/K_{max}$, the ratio of minimum to maximum kinetic energy  as a measure characterizing a particular wave regime as standing or \ADD{propagating} waves. Similar to $\left\langle K\right\rangle$, this quantity is defined for the saturated regime. For purely standing waves, $R= 0$, while \ADD{$R\rightarrow 1$} for purely  \ADD{propagating} waves with vanishingly thin wave beams \ADD{(in the linear regime with a vanishingly small viscosity)}. In realistic systems with wave beams of finite thickness, we observe $0 <R < 1$. 

\begin{figure}
	\begin{centering}
		   \includegraphics[width=0.85\textwidth]{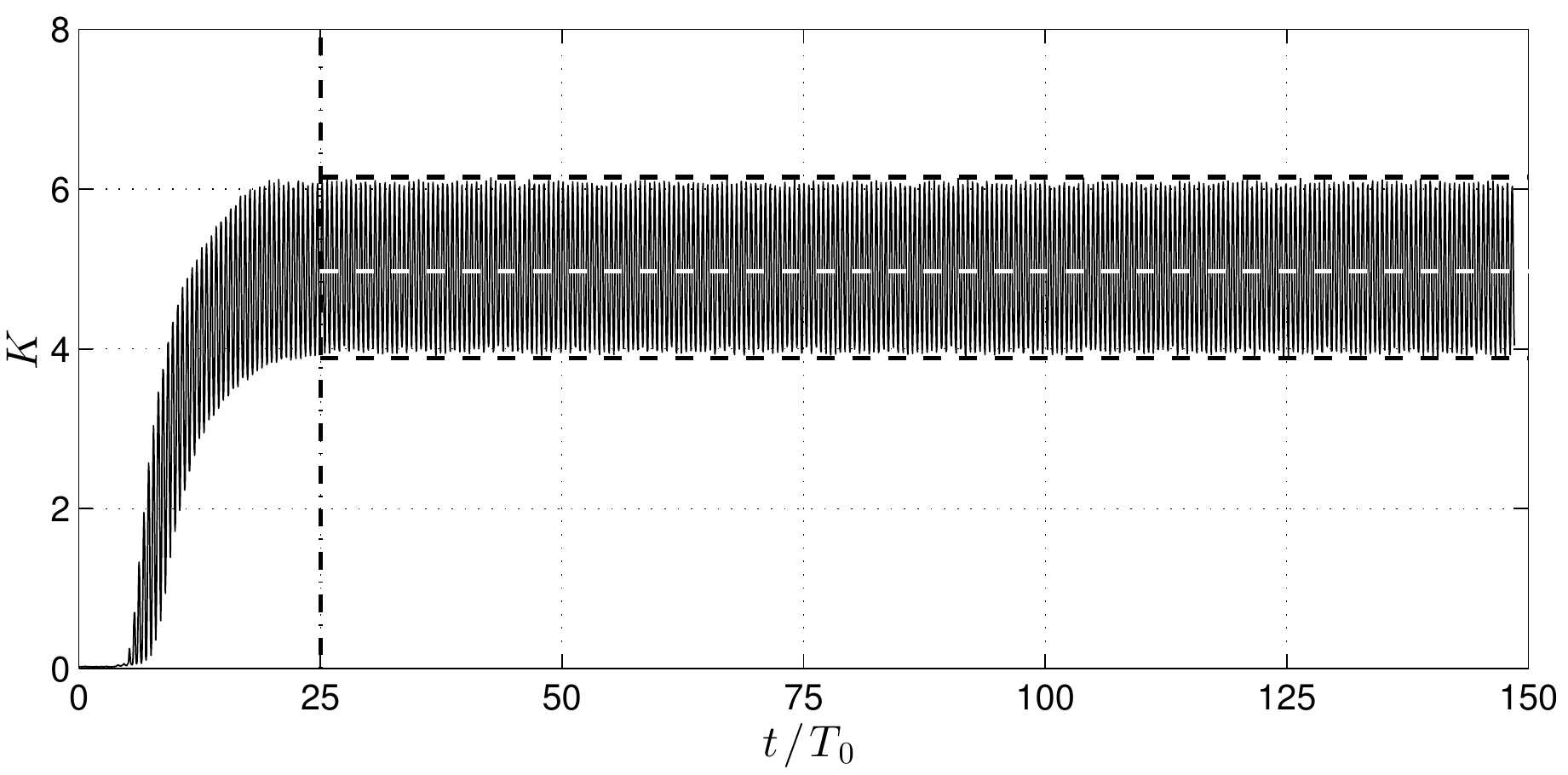}
		\caption{Typical experimental time-history of the nondimensional kinetic energy. The vertical dotted dashed line indicates the instant where the kinetic energy saturates. The average of the signal, computed from this instant and to the end of the experiment, is represented by the dashed white line. The two horizontal dashed black lines represent $K_{min}$ and $K_{max}$, the minimal and maximal values bounding the oscillations of the kinetic energy during the saturated state. {The parameters for this experiment are: $\Omega_0$=0.59, $H$=300 mm, $L$=450 mm, $\alpha=27.3^\circ$, $\tau=1.84$, $d=0.31$ and $a=1.5$ mm. For this experiment, $\left\langle K\right\rangle=5$ and $R=0.65$.}\label{kinetic_energy_vs_time}}
	\end{centering}
\end{figure}

\begin{table}
\begin{center}
\begin{tabular}{l|c|c|c|c|c|c|c|c|c|c|c}
 \#&$\alpha $ & $\Omega_0$ & $d$ & $\tau$  &\null\hskip 2truecm\null&  \#&$\alpha $ & $\Omega_0$ & $d$ & $\tau$ \\
1 &11.3 & 0.552 & 0.73 & 1.93  &\hskip 1truecm & 26&31.2 & 0.719 & 0.20 & 1.24 \\
2&11.3 & 0.554 & 0.73 & 1.90  & & 27&31.2 & 0.731 & 0.20 & 1.22 \\
3&11.3 & 0.574 & 0.73 & 1.86  & & 38 &34.6 & 0.530 & 0.09 & 1.93 \\
4&11.3 & 0.588 & 0.73 &  1.83 & & 39 & 34.6 & 0.545 & 0.09 & 1.85\\
5&11.3 & 0.602 & 0.73 & 1.81  & & 30 & 34.6 & 0.588 & 0.09 & 1.61 \\
6&11.3 & 0.609  & 0.73 & 1.79 & & 31 & 34.6 & 0.643 & 0.09 & 1.43 \\
7&24.2 & 0.551 & 0.41 & 1.94  & & 32& 34.6 & 0.643 & 0.09 &  1.41 \\
8&24.2 & 0.559 & 0.41 & 1.89  & & 33 & 34.6 & 0.669 & 0.09 & 1.34 \\
9&24.2 & 0.587 & 0.41 & 1.85  & & 34& 34.6 & 0.707 & 0.09 & 1.15 \\
10&24.2 & 0.615 & 0.41 &  1.74 & & 35& 37.5 & 0.673 & 0.00 & 1.85 \\
11&24.2 & 0.642 & 0.41 &  1.42 & & 36& 37.5 & 0.707 & 0.00 & 1.78 \\
12&24.2 & 0.668 & 0.41 & 1.37  & & 37& 37.5 & 0.743 & 0.00 & 1.42 \\
13&28.5 & 0.554 & 0.28 & 1.90  & & 38& 37.5 & 0.755 & 0.00 & 1.30 \\
14&28.5 & 0.574 & 0.28 & 1.83 & & 39& 37.5 & 0.766 & 0.00 & 1.04 \\
15&28.5 & 0.602 & 0.28 & 1.71 & & 40 &40.3 & 0.552 & -0.10 & 1.88 \\
16&28.5 & 0.622 & 0.28 & 1.62 & & 41 & 40.3 & 0.559 & -0.10 & 1.85 \\
17&28.5 & 0.656 & 0.28 & 1.49 & & 42& 40.3 & 0.602 & -0.10 & 1.55 \\
18&28.5 & 0.695 & 0.28 & 1.32  &   & 43 &40.3 & 0.616 & -0.10 & 1.50 \\
19&28.5 & 0.707 & 0.28 &  1.30 &   & 44 &40.3 & 0.643 & -0.10 & 1.40 \\
20&31.2 & 0.554 & 0.20 & 1.90 & & 45 &40.3 & 0.731 & -0.10 & 1.22 \\
21&31.2 & 0.574 & 0.20 & 1.83 & & 46 &45.0 & 0.530 & -0.30 & 1.86 \\
22&31.2 & 0.602 & 0.20 & 1.68 & & 47 &45.0 & 0.559 & -0.30 & 1.85 \\
23&31.2 & 0.636 & 0.20 & 1.54 & & 48 &45.0 & 0.602 & -0.30 & 1.50 \\
24&31.2 & 0.669 & 0.20 & 1.42 & & 49 &45.0 & 0.629 & -0.30 & 1.40 \\
25&31.2 & 0.707 & 0.20 & 1.24 & & 50 &45.0 & 0.669 & -0.30 & 1.36 \\
\end{tabular}
\caption{Parameters for the series of experiments represented in figure~\ref{diagram} and
studied in this paper.
For all these experiments, the working depth is {300}~mm, the working bottom length is {455$\pm$10}~mm,
and the amplitude of the generator is $a={15}$~mm.  $\#$ is the number of the experiment (from 1 to 50),
$\alpha$ the angle (in $\circ$) of the slope, $\Omega_0$ the nondimensional forcing frequency, $d$ and $\tau$
the corresponding nondimensional parameters of the diagram presented in figure~\ref{nouveaudiagram}.}
\label{tabular:shorttermparameters}
\end{center}
\end{table}

Using $\langle K\rangle$ and $R$ as variables, we performed a series of 50 {\it short-term} experiments (parameters are given in Table~\ref{tabular:shorttermparameters}) with stable $(1,1)$ attractors to study the structure of their domain of existence, the so-called Arnold tongue, as a function of the two parameters~$(d,\tau)$ controlling the convergence of wave rays.  For the particular case of trapezoidal geometry, \cite{MBSL1997} introduce $d=1-(2H/L) \tan \alpha$ as the control parameter of the slope 
 and $\tau=(2H/L) \sqrt{1/\Omega_0^2-1}$ as the control parameter of {the wave-ray pattern}. The limiting case of triangle geometry ($d=-1$) is typically characterized by the presence of a point attractor at a vertex of the triangle, while the case of rectangular  geometry ($d=1$) corresponds to classic normal modes~\citep{McEwan1971} for a discrete set of $\tau$. Originally,  $(d, \tau)$ diagrams have been suggested in~\cite{MBSL1997} as a highly convenient tool for the classification of the inviscid wave regimes in a trapezoidal domain filled with a linearly stratified fluid. Lyapunov exponents have been used as a variable characterizing the convergence of wave rays toward limit cycles. The plot of the Lyapunov exponents as a function of $(d, \tau)$ is shown in figures \ref{nouveaudiagram} and~\ref{diagram} in greyscale. The white (respectively grey) domains correspond to strong (respectively weak) convergence \ADD{onto wave attractors}. 
Wave attractors with high rate of convergence toward a simple well-defined structure are found in white regions (as anticipated, the so-called Arnold tongues). The dark- and light-grey regions have a highly complicated structure where the Lyapunov exponent varies in a wide range, occasionally reaching nearly-zero values (quasi-resonances). Ray tracing reveals for these regions a slow convergence toward attractors of high complexity \ADD{with a high number of boundary reflections}.  Identically zero Lyapunov exponents are found at a discrete set of lines corresponding to global resonances (seiche modes).

It can be seen that the domain of existence of $(1,1)$ attractor in figure \ref{diagram} has a triangular shape. \ADD{In a linear viscous regime at fixed values of $Ar$ and $H/L$, the quantities $\left\langle K\right\rangle$ and $R$ are the functions of a coordinate in $(d,\tau)$ plane.}
Combining the information presented in terms of~$\left\langle K \right\rangle$ and~$R$, we can conclude that there are two distinct regions with  different regimes for $R$:
\begin{itemize}
\item[i)] the central part of the Arnold tongue with high values 
of~$R$: it is indicating a \ADD{propagating} wave system (denoted NP in the remainder of the text)
corresponding to a classic case of wave attractor with thin well-defined branches. {It is clearly seen in the left panel of figure~\ref{diagram} that there exists an optimum range in $(d,\tau)$-space, \ADD{where the mean kinetic energy of a well-focused attractor is maximized}. As the focusing increases (negative $d$), the excitation of high-energy attractors is hindered by increased dissipation in narrow wave beams.} 
\item[ii)]  the border regions of the Arnold tongue are typically characterised by low values of~$R$ {due to geometric degeneration of the attractor}. In that domain referred as NS, waves are nearly standing \ADD{and the mean kinetic energy of attractor is low except,} the right tip of the Arnold tongue, \ADD{where} we observed high values of~$\left\langle K \right\rangle$ and low values of $R$, typical features for standing waves. Indeed, the right \ADD{tip} of the Arnold tongue corresponds to  \ADD{$d=1$ and $\tau=2$}, i.e.  \ADD{to an exact standing mode. Thus, the secondary maximum of $\left\langle K \right\rangle$ in the vicinity of the right corner of the Arnold tongue is associated with optimum condition of excitation of a nearly standing wave in a nearly rectangular domain.} 
\end{itemize}

\begin{figure}
   \begin{centering}
   \includegraphics[width=0.95\textwidth]{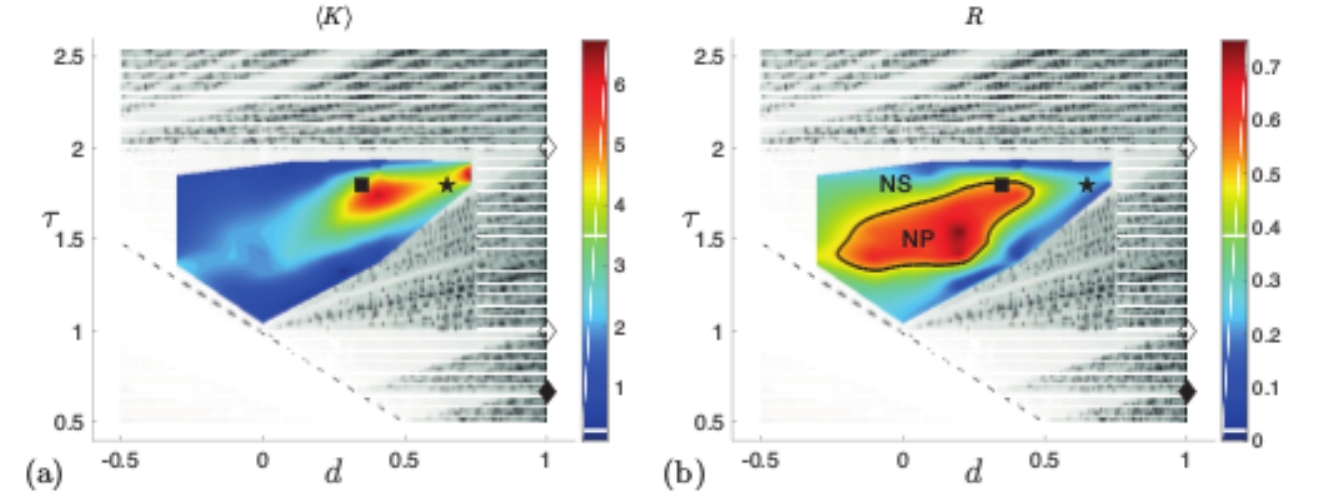}
   \caption{Structure of Arnold tongue of a stable $(1,1)$ attractor in terms of the mean normalized kinetic energy~$\left\langle K\right\rangle$ (panel a) and the ratio of minimum to maximum kinetic energy~$R$ (panel b). The results are plotted from the 50 experiments described in Table~\ref{tabular:shorttermparameters}. Superimposed on this picture are three different symbols corresponding to the location of the three main long-term experiments discussed throughout this paper:   
   $\blacklozenge$ and $\lozenge$ for the case of the rectangular domain discussed by~\cite{McEwan1971}  (see \S~\ref{sectionMcEwan}), $\blacksquare$  for the well-focused $(1,1)$ attractor discussed in~\cite{BrouzetEPL2016} with propagating waves (see \S~\ref{TRIpropagating}),  $\bigstar$ for {the weakly-focused $(1,1)$ attractor described} here (see \S~\ref{TRIstanding}). {The solid black line shows the contour at the value $R=0.5$, enclosing the region with nearly propagating waves (NP), while nearly standing waves (NS) are outside this region.}
   	\label{diagram}}
\end{centering}
\end{figure}

\section{Three scenarios for the onset of triadic resonance in a trapezoidal domain}

In this section, we discuss how the choice of the operating point in the Arnold tongue of $(1,1)$ attractors affects the observed scenario for the onset of triadic resonance instability~(TRI). {In Table~\ref{tabular:parametersMainexperiments} are specified the experimental conditions 
of the three main {\it long-term} experiments represented by the symbols $\blacklozenge$, $\blacksquare$  and $\bigstar$ in figure~\ref{diagram}.}

\begin{table}
\begin{center}
\begin{tabular}{l|c|c|c|c|c|}
Reference& Unit & \cite{McEwan1971} & \cite{BrouzetEPL2016} & Present paper (2016)\\
\hline
Section & \S & \ref{sectionMcEwan} & \ref{TRIpropagating} & \ref{TRIstanding}\\
Symbol  in Fig.~\ref{diagram} & --& $\blacklozenge$ & $\blacksquare$ & $\bigstar$\\
$H$ &mm & 326 & 303 & 303\\
$L$ &mm & 377 &444& 444\\
$a$ & mm& 4.8& 5 {or 10}  &5 {or 10}  \\
$\alpha $ & $\circ$ &  90&25.4&14.3\\
$\Omega_0$ &--& 0.655 & 0.61&0.60 \\
$d$ &--& 1 & {0.35}&0.65\\
$\tau$ &--& 0.66 & {1.80}&1.80\\
$\Omega_1$ &--& 0.397& 0.36&0.33\\
$\Omega_2$ &--& 0.277 & 0.25&0.27\\
$|\ell_*|$ &\ADD{rad$\cdot$}m$^{-1}$ & 25.0& 7.8 &7.8\\
$|m_*|$ &\ADD{rad$\cdot$}m$^{-1}$ &  9.6& 10.4 &10.4\\
$|\ell_0|$ &\ADD{rad$\cdot$}m$^{-1}$ &  25.0&33&9.5\\
$|m_0|$ &\ADD{rad$\cdot$}m$^{-1}$ &  9.6&37& 12.6\\
$|\ell_1|$ &\ADD{rad$\cdot$}m$^{-1}$ &  50.0 & 80& 25.3\\
$|m_1|$ &\ADD{rad$\cdot$}m$^{-1}$ & 38.5 & 200&66.4\\
$|\ell_2|$ &\ADD{rad$\cdot$}m$^{-1}$ & 25.0 & 45 &15.8\\
$|m_2|$ &\ADD{rad$\cdot$}m$^{-1}$ & 28.9 & 170&53.8
\end{tabular}
\end{center}
\caption{Parameters for the three main and long term experiments discussed in this paper. $\ell_*$ and $m_*$ are respectively the horizontal and vertical forcing wavenumbers.
The horizontal and vertical wave vector components are denoted ($\ell_i,m_i)$ in which the subscript  $i$ refers to the primary ($i=0$) or to the secondary waves ($i=1$ and 2). $\Omega_i$ is the corresponding nondimensionalized frequency. In \cite{McEwan1971}, we have chosen the data with the frequency triplet that is close to those considered in \cite{BrouzetEPL2016} and in the present paper. The values of wave vectors are evaluated from the parameters of standing waves presented in \cite{McEwan1971}.  }
\label{tabular:parametersMainexperiments}
\end{table}

\subsection{{TRI} with purely standing waves: the degenerate case of the rectangular domain}\label{sectionMcEwan}

The case of TRI in standing waves in a rectangular domain has been studied in~\cite{McEwan1971}, represented by the lozenges
$\blacklozenge$ and $\lozenge$ in figure~\ref{diagram}. These cases fall on the vertical line $d = 1$ where $R=0$. The theoretical analysis presented in~\cite{McEwan1971} assumes that standing primary wave oscillating at frequency $\Omega_0$ can feed two standing secondary waves oscillating at frequencies $\Omega_1$ and $\Omega_2$ provided spatial and temporal conditions of the triadic resonance are satisfied and provided the damping in the system is below a given threshold. The damping is supposed to be primarily associated with the loss of energy in boundary layers at the \ADD{side} walls of the domain. The temporal resonance condition of the TRI amounts to $\Omega_0=\Omega_1 \pm \Omega_2$. Since all the waves involved into triadic resonance are standing ones, their half-lengths in horizontal and vertical directions should be equal to $L/M$ and $H/N$ where $M$ and $N$ are integer numbers. Then, the spatial resonance condition can be written as $M_0=M_1 \pm M_2$ and $N_0=N_1 \pm N_2$. In practice, high energetic contents are encountered for the triads with ``plus" and ``minus" signs in temporal and spatial resonance conditions, respectively, i.e. with the cases where secondary waves oscillate at a smaller frequency than the primary wave, and the energy is transferred to shorter spatial scales, privileging the direct cascade. All three waves ($j=0,1,2$) should satisfy the dispersion relation: $(N_{j}/M_{j})^2 (L/H)^2=(1-\Omega_{j}^2)/\Omega_{j}^2$. 

In the experiments described in~\cite{McEwan1971}, the forcing typically corresponds to low mode with $N_0=1$ and $M_0$ varying from 1 to 5. It has been shown experimentally that, at the laboratory scale, it is possible to transfer energy to the waves that are a few times shorter than the primary one. For example, for the low-mode forcing at $N_0=1$ and $M_0=3$, \cite{McEwan1971} observed secondary waves with $N_1=\ADD{4}$, $M_1=\ADD{6}$ and $N_2=3$, $M_2=3$. 

\subsection{{TRI} with propagating waves: the case of well-focused attractor\label{TRIpropagating}}  

This case of TRI has been described in~\cite{SED2013} and~\cite{Brouzetetal2016}. It is a typical scenario of instability in the central region of the Arnold tongue corresponding to $(1,1)$ attractor in $(d, \tau)$ diagram. An experiment reported in~\cite{BrouzetEPL2016}, also located in the central region of the $(1,1)$ Arnold tongue, shows a similar instability, illustrated on figure~\ref{figurearajouter}. This experiment is represented  by the symbol $\blacksquare$ in figure~\ref{diagram} for $a=5$~mm. For this regime, $\langle K\rangle$ is large and {$R \gtrsim 0.5$}. Accordingly, the $(1,1)$ attractor has thin well-localised branches. \cite{SED2013} and~\cite{Brouzetetal2016} have shown that the onset of instability in such a wave attractor occurs  locally first in the most energetic wave beam which has the largest amplitude but the smallest width. This scenario of instability is consistent with the one discussed in~\cite{BDJO2013,BSDBOJ2014} and~\cite{KarimiAkylas2014}. This is also what is observed in~\cite{BrouzetEPL2016}, at the very beginning of the triadic cascade, as shown in figure~\ref{figurearajouter}.

Owing to local character of instability \citep{SED2013,BrouzetEPL2016,Brouzetetal2016}, the confinement of fluid domain does not affect the instability directly. The focusing provides the energy transfer from the input perturbation which has the scale of vertical size of the fluid domain to the scale associated with the width of the attractor beams, which serves as a primary wave.  Subsequent triadic resonance transfers energy to secondary waves. The overall efficiency of the energy transfer from large- to small-scale motions is remarkably high, even at the laboratory scale. \cite{SED2013} describe the case where the typical wavelength of the primary wave represented by the beam of the attractor is roughly 9 times shorter than the wavelength of the input forcing, and one of the secondary waves has even a wavelength roughly 3 times shorter than the primary wave, leading to a reduction factor of 25. This is {also visible in figure~\ref{figurearajouter} where the primary and the two secondary waves have been disentangled {using the Hilbert transform~\citep{MGD2008}}.

\begin{figure}
   \begin{centering}
 \includegraphics[width=\textwidth]{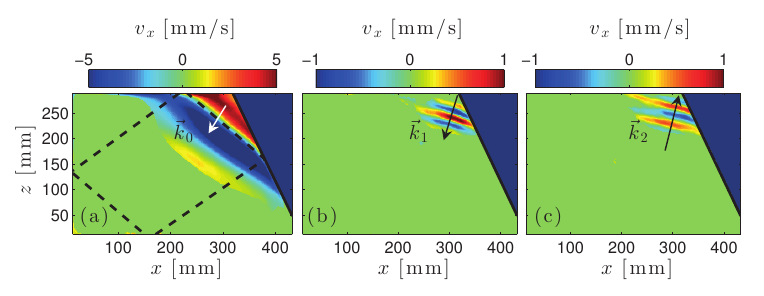}
   \caption{
   Horizontal velocity fields {for the experiment indicated by the symbol $\blacksquare$ in figure~\ref{diagram}
    and Table~\ref{tabular:parametersMainexperiments} for $a=5$ mm,} filtered at $\Omega_0$ (panel a), $\Omega_1$ (panel b) and $\Omega_2$ (panel c) in frequency and in space to keep only the relevant directions. The velocity field is displayed only where the wave amplitude is larger than $15\%$ of the maximum. The wave vectors are represented by the arrows. Black dashed lines show the billiard geometric prediction of the attractor. \label{figurearajouter}}
\end{centering}
\end{figure}

\subsection{{TRI} with nearly standing waves: the case of weakly focused attractor.} 
\label{TRIstanding}

In this subsection, we describe a new scenario for the onset of instability which is intermediate between those described in \S 4.1. and \S 4.2. This new scenario is typical for weak focusing. 
The observation of triadic resonance has been performed for $(d,\tau)=(0.65,1.80)$ and $a~=~5$~mm. This experiment
is represented  by the symbol $\bigstar$ in figure~\ref{diagram}.

\subsubsection{The onset of instability}
The snapshot of wave pattern after $200$ periods of oscillations (approximately $30$ minutes of observations) is shown in figure \ref{20140619_snapshot}(a). The time history of oscillations measured at the point~A, defined  in figure~\ref{20140619_snapshot}(a), is presented in figure~\ref{20140619_snapshot}(b). After a rather short transient of approximately $25$ periods, the signal is stable: the attractor is set. Then, one can clearly see the apparent instability developing after $400$ forcing periods, i.e. after more than $1$~hour of observations.

\begin{figure}
   \begin{centering}
  \includegraphics[width=0.8\textwidth]{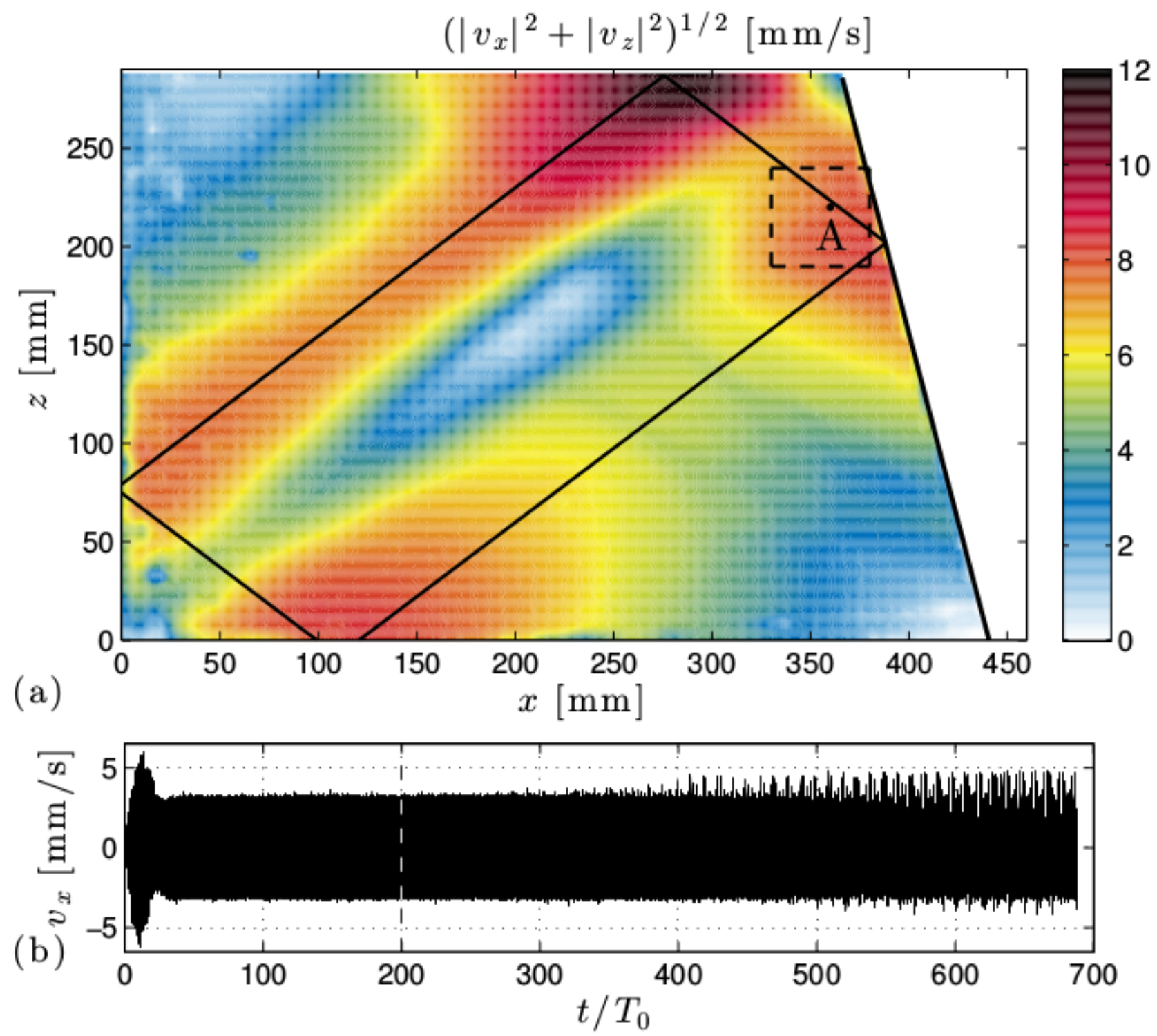}
   \caption{Velocity field for the experiment indicated by the symbol $\bigstar$ in figure~\ref{diagram}
  {with a=5 mm}  (see Table~\ref{tabular:parametersMainexperiments} for the parameters).
(a) Picture of internal wave field, filtered around $\Omega_0$, $200~T_0$ after the start of the wave maker and black lines show the billiard geometric prediction of the attractor. The picture presents the absolute value of the norm of the velocity vectors. The large value around the upper corner of the attractor is due to the presence of a small homogenous layer above the stratified domain. (b) the time-history of oscillations at the point A shown in (a). The vertical dashed line in (b) shows the instant of the picture (a) while the dashed square in (a) shows the region where the time-frequency diagram in figure \ref{20140619_tps_freq} has been calculated. The initial transient is due to the setting of the attractor, while the instability is noticeable after roughly 400 periods of oscillations. 
   \label{20140619_snapshot}}
\end{centering}
\end{figure}

The development of the spectrum of wave motion over time is presented in figure \ref{20140619_tps_freq}(a) with the use of the time-frequency diagram. It is calculated at each point in space, as in~\cite{BDJO2013}, with the formula 
\begin{equation}
S_r(\Omega,t)=\left\langle \left| \int_{-\infty}^{+\infty} \! v_r(x,z,\tau)e^{i\Omega N\tau}h(t-\tau)\, \mbox{d}\tau 
\right|^2\right\rangle_{xz},
\end{equation}
where $h$ is an Hamming window and $v_r$ the component of the velocity field, $r=x$ or $z$. The calculations are performed with the Matlab toolbox described in~\cite{Flandrin1999}.

\begin{figure}
   \begin{centering}
	\includegraphics[width=0.8\textwidth]{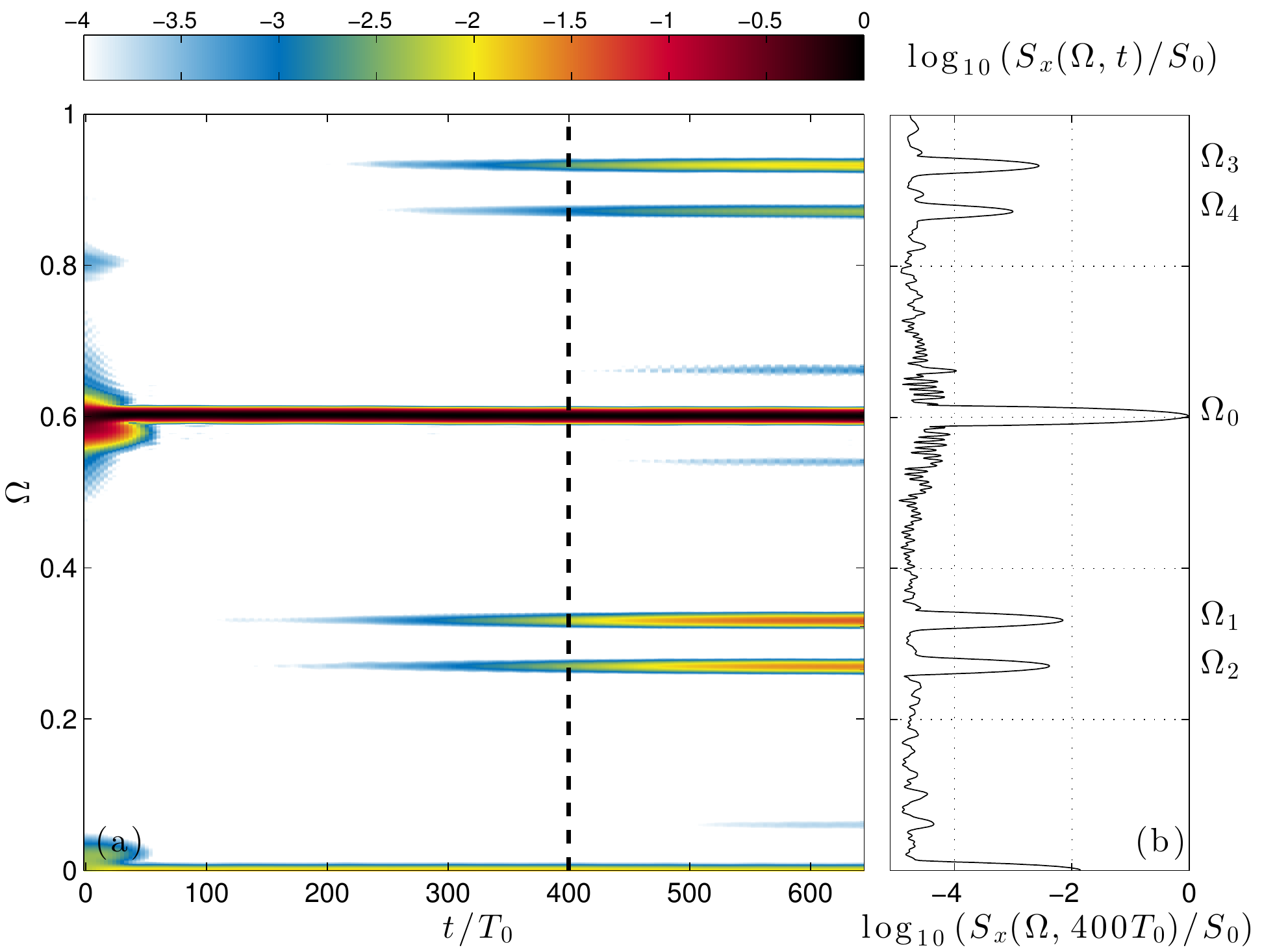}
   \caption{{\em Weak TRI}. (a) Time-frequency diagram of internal-wave field, obtained from a $50$~mm side square region, surrounding the point A shown in figure \ref{20140619_snapshot}(a);  (b) vertical cut of the diagram along the frequency axis at $t=400T_0$, indicated by the white dashed line. $S_0$ corresponds to the time average of the frequency component associated with the primary wave $\Omega_0$.  Results are presented for the experiment indicated by the symbol $\bigstar$ in figure~\ref{diagram}
    (see Table~\ref{tabular:parametersMainexperiments} for parameters).     \label{20140619_tps_freq}}
\end{centering}
\end{figure}

The appropriate choice of the length of the Hamming window allows to tune the resolution in frequency and time. Typically, the data are averaged over a finite analysing area which, as a limit, can extend over the whole fluid domain. Figure \ref{20140619_tps_freq}(a) presents the time-frequency diagram for the case illustrated in figure \ref{20140619_snapshot}. The diagram is calculated for the data averaged over the rectangle shown in figure \ref{20140619_snapshot}(a) around point A. It can be seen that, initially, the signal is entirely dominated
 by the forcing frequency $\Omega_{0}=0.60$ corresponding to the primary (carrier) wave. Oscillations with frequencies $\Omega_{1}=0.33$ and $\Omega_{2}=0.27$ slowly develop with time. These frequencies correspond to two secondary waves generated by TRI. They satisfy the frequency conditions for the triadic resonance:
\begin{equation}
\Omega_{1}+\Omega_{2}=\Omega_{0}.
\label{eq:TRI}
\end{equation}
In addition, one can also see two peaks satisfying differential conditions:
\begin{equation}
\Omega_{3}-\Omega_{1}=\Omega_{0} \qquad\mbox{and}\qquad
\Omega_{4}-\Omega_{2}=\Omega_{0}.
\end{equation}

\subsubsection{The components of the wave pattern}

\begin{figure}
   \begin{centering}
\includegraphics[width=0.85\textwidth]{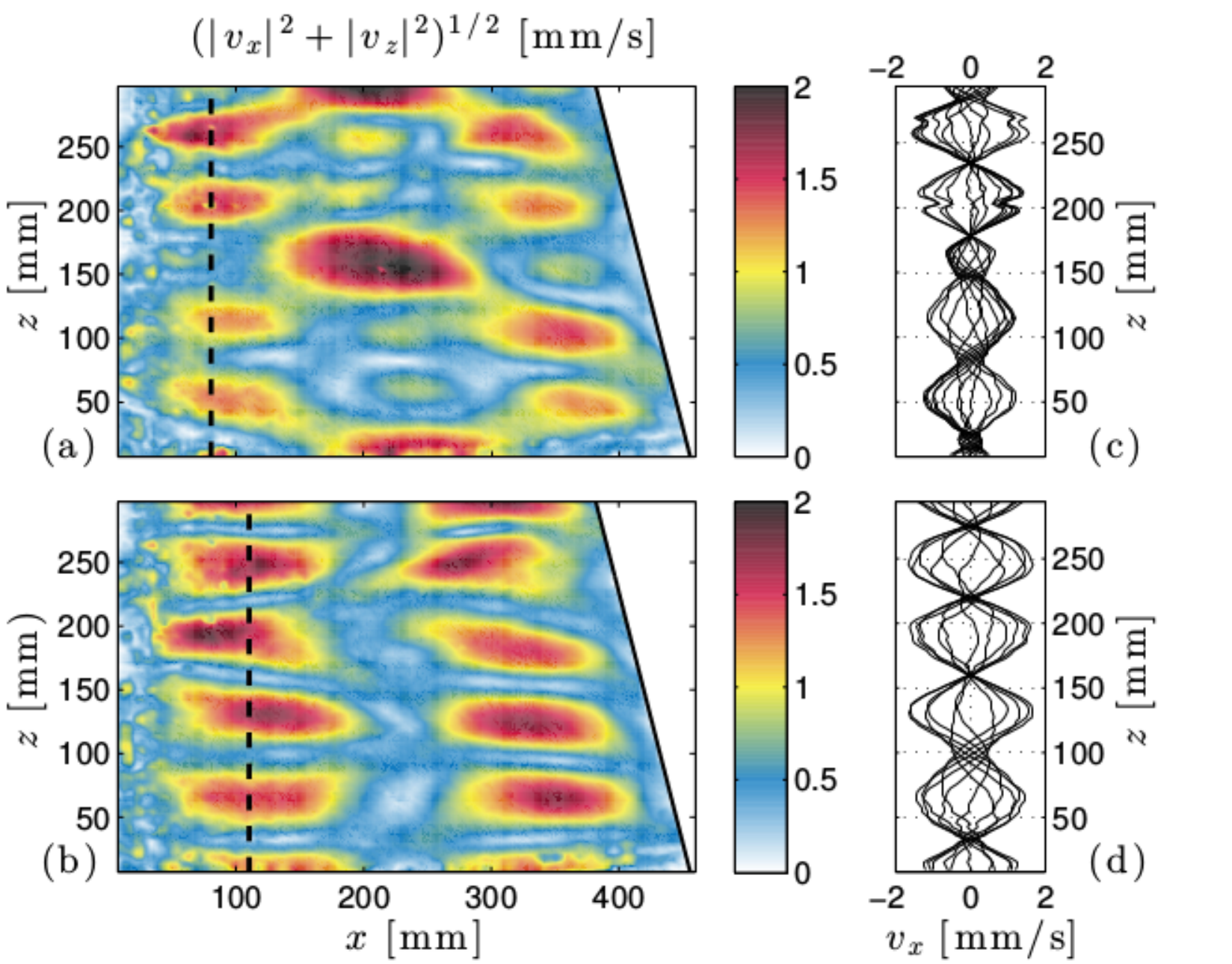}
   \caption{Components of the wave field corresponding to frequencies $\Omega_{1}$ (panel a) and $\Omega_{2}$ (panel b), obtained with the Hilbert transform around $620~T_0$ using a time window $85~T_0$ long. Results are presented for the experiment indicated by the symbol $\bigstar$ in figure~\ref{diagram} with a=5 mm
    (see Table~\ref{tabular:parametersMainexperiments} for the parameters). Note that the components oscillating at $\Omega_{1}$ and $\Omega_{2}$  correspond to {nearly} standing waves as is clearly seen on the sequences of wave profiles shown on the right of each picture.     \label{20140619_filles}}
\end{centering}
\end{figure}

Figure \ref{20140619_filles} shows the decomposition of the wave field into the components corresponding to the secondary waves \ADD{while the primary wave has already been exhibited in figure~\ref{20140619_snapshot}(a).} The decomposition, performed with the help of Hilbert transform~\citep{MGD2008},
reveals node-antinode patterns of amplitudes of secondary waves, similar to Chladni figures. These patterns are present soon in the experiment, but their amplitude increases gradually. Thus, this is very different from TRI presented in \cite{SED2013} where the waves are created on a very localized region of the tank. Here, the secondary waves appear directly everywhere in the tank, in a shape presented in figure~\ref{20140619_filles}. The sequences of wave profiles measured along the vertical line indicated in the images are also shown in the right panels. It can be seen that the secondary-wave motion is represented by standing waves of high vertical modes. \ADD{They do not correspond to global eigenmodes of the trapezoidal basin (represented by the black lines of the
 $(d,\tau)$ diagram) but to quasi-global resonances (dark regions of the $(d,\tau)$ diagram, as shown in figure~\ref{nouveaudiagram}(c). Diffusion induces a certain thickness to the branches and thus allows the secondary waves to be quasi-standing and to correspond to quasi-global modes of the trapezoidal basin.}

The standing wave patterns can be further decomposed into the sums of propagating waves {whose wave vectors can be measured} (see data in Table 2). All vectors satisfy the dispersion relation individually. The length of the primary wave can be estimated~\ADD{\citep{MGD2008}} as $\lambda_0=2\pi/|\mathbf{k}_{0}|=39.8~$cm with horizontal and vertical components $\lambda_{x0}=66.1~$cm and $\lambda_{z0}=49.7~$cm. The quantities $\lambda_{x0}/2$ and $\lambda_{z0}/2$  {are reasonably close to} the distances from the left top corner of the trapezoidal fluid domain to the reflection points of the attractor. The ray tracing yields the distances to the reflection points 28 and 21.6 cm in horizontal and vertical directions, respectively. Therefore, in the present case, the primary wave has a length scale associated with the global geometry of attractor. \ADD{The length scale associated with the beam width does not clearly manifest itself. Indeed, weakly focused attractors with elongated shape have two parallel beams which can interact with each other since the width of beams and the inter-beam distance are comparable quantities. As result the notion of the beam width is difficult to define for such attractors, in contrast with spatially localised beams of well-focused attractors.}  \ADD{In the latter case} described in~\cite{SED2013} (see \S~\ref{TRIpropagating} and symbol $\blacksquare$ in figure~\ref{diagram}), we have seen that the length of the primary wave is associated with the width of the attractor \ADD{beam}. 

\subsubsection{The growth rate}

Horizontal cuts in figure \ref{20140619_tps_freq} at  
$\Omega_1$ and $\Omega_2$
are plotted in figure \ref{20140619_bispectre_coupe}. At low amplitude of secondary waves, the linear trend is obscured by the instrumental noise, while toward the end of the experiment there is a  trend to saturation of amplitudes of secondary waves. The vertical logarithmic representation emphasizes a remarkably linear growth for about three decades (!) in the {magnitude of the spectral peaks corresponding to} secondary waves.
 The growth rate of the instability can be estimated to be around $\sigma = 6.5\times10^{-4}$~s$^{-1}$, which leads to a characteristic growth time of the instability around $1500$~s (i.e. half an hour {or $150 T_0$}). \ADD{The theoretical value for monochromatic plane wave, taking account viscosity~\citep{KoudellaStaquet2006,BDJO2013} is equal to $2.1\times10^{-2}$~s$^{-1}$, that is two order of magnitudes larger than the experimental value. The difference is due to the spatial confinement of the attractor, which differs clearly from monochromatic plane wave, as underlined by~\cite{KarimiAkylas2014} and \cite{BSDBOJ2014}.}

\begin{figure}
	\begin{centering}
	\includegraphics[width=0.5\textwidth]{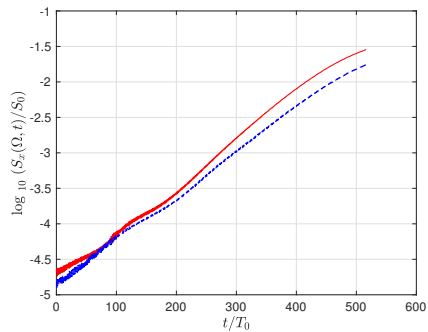}
		\caption{Cuts along the time-axis of the time-frequency diagram presented in figure~\ref{20140619_tps_freq} at frequencies $\Omega_{1}$ (solid red line) and $\Omega_{2}$ (dashed blue line).
		\label{20140619_bispectre_coupe}}
	\end{centering}
\end{figure}

Th\ADD{e} \ADD{instability growth} time has to be compared with the time scale~$\tau_w$ set by the limited size of the test tank. It can be defined as the time for secondary waves to perform a horizontal return-trip in the tank: $\tau_w=2L/c_{gx}$, where $c_{gx}$ is the group velocity in the horizontal direction. This group velocity can be measured using the wave numbers of the secondary waves estimated via Hilbert transform and wave number filtering. It gives $\tau_w\approx 60~$s, which is very small in comparison with the instability growth time of $1500~$s. It is worth to note that the horizontal group velocity is a good measure of the return-trip time of secondary waves because these waves are only slightly inclined. Indeed, their propagation angle, with respect to the horizontal direction, is close to $20^{\circ}$ since $\Omega_1$ and $\Omega_2$ are around $\Omega_0/2=0.3$.

The data presented in~\cite{SED2013} or in figure~\ref{figurearajouter} clearly shows that the triadic resonance develops first in the most energetic branch of the wave attractor, the one connecting the inclined slope to the surface. \ADD{The instability growth time is of $30$~s.}
The data on the wave-vector components presented in~\cite{SED2013} yield $\tau_w$ around $180$ and $320~$s for the two secondary waves involved into the triplet which is measured at time around $300~$s after emergence of detectable secondary waves. Therefore, the secondary waves could make only between 1 and 2 return trips, and owing to the presence of the vertical component of the group velocity they could not return to their generation site. So, the onset of the instability described in~\cite{SED2013}, and also seen in the experiment represented by the $\blacksquare$ on the $(d,\tau)$ diagram is to a good approximation a spatially isolated {\it local} event, while the onset and subsequent development of instability described in the present Section is a {\it global} event, which is inherently related to the confinement. This event has many common features with the instability in a rectangular domain described in~\cite{McEwan1971}. 

\section{Well-developed TRI cascade in well-focused attractor: presence of secondary standing waves}

In the previous Section, we have demonstrated the effect of the ratio between the typical growth time of the primary instability $\tau_{*}$ and the duration of the horizontal return-trip $\tau_w$ of secondary waves in the tank. If  $\tau_{*} \ll \tau_w$, the onset of instability is local, while for $\tau_{*} \gg \tau_w$ the secondary waves are more likely to appear in form of standing waves. However, $\tau_{*}$ and $\tau_w$ do not form an exhaustive set of time scales. The evolution of the system to saturated state where the energy injection is in balance with dissipation may have a very long time-scale $\tau_{sat}$. Over the time span $0 < t < \tau_{sat}$, the system may experience a cascade of triadic resonance instabilities generating new time-scales of the return trips $\tau_w$. For the full study of dynamics, the system should be observed during the time $\tau_{obs} > \tau_{sat} \gg \tau_{w}$. Thus, long-term experiments with $\tau_{obs} \gg \tau_{w}$ can reveal the effects of confinement for the system with  $\tau_{*} \ll \tau_w$ similar to the one described in \cite{SED2013}. In this section, we analyse the long-term evolution of the spectrum in the case with strong focusing 
($\blacksquare$ on $(d,\tau)$ diagram) and present  for the first time
the evidence of the effects due to confinement.

\subsection{Time-frequency diagram}
\begin{figure}
   \begin{centering}
	\includegraphics[width=0.8\textwidth]{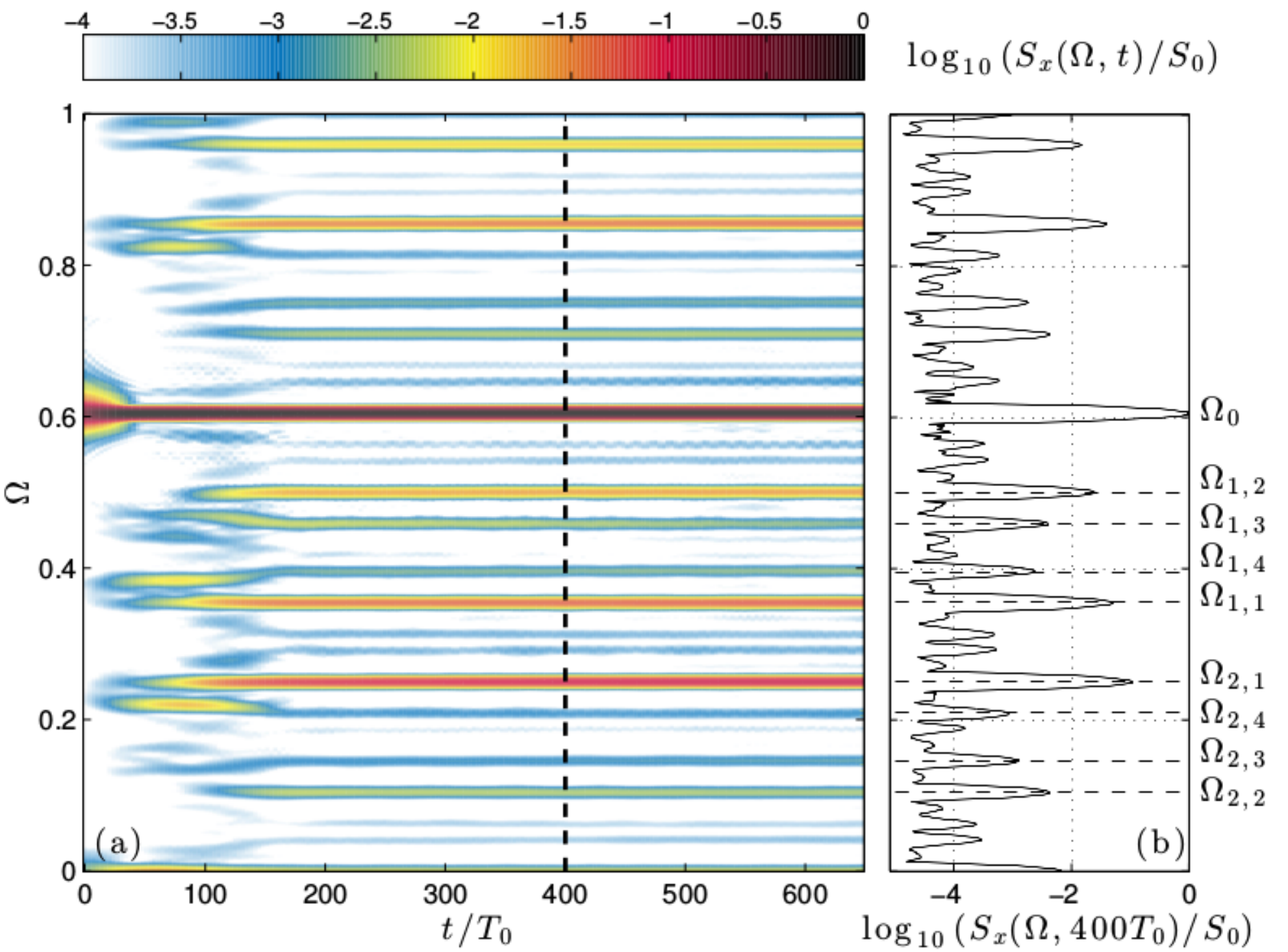}
   \caption{{\em TRI cascade}. (a) Time-frequency diagram of internal-wave field, obtained from a $5\times5$~cm$^2$ side square region {in the most energetic wave beam};  (b) Vertical cut of the diagram along the frequency axis at $t=400T_0$, indicated by the black dashed line. $S_0$ corresponds to the time average of the frequency component associated with the primary wave $\Omega_0$. The different frequency peaks corresponding to the four first couples are labelled. Results are presented for the experiment indicated by the symbol $\blacksquare$ in figure~\ref{diagram} {with a=5 mm}
    (see Table~\ref{tabular:parametersMainexperiments} for the parameters). 
     \label{20140513_tps_freq}}
\end{centering}
\end{figure}

The time-frequency diagram  {of the long-term experiment with well-focused attractor ($\blacksquare$) at $a=5$~mm} is presented in figure~\ref{20140513_tps_freq}. One can identify at least eight couples of secondary waves in panel~(a). The four most energetic couples in the spectrum are labelled in panel~(b). The secondary frequencies are named using two indices as follows. The first index $i$ indicates the position in the couple, $i=1$ for frequencies higher than $\Omega_0/2$ or $i=2$ for frequencies smaller than $\Omega_0/2$. The second index $n$ corresponds to the number of the couple. Thus, $n$ varies from $1$ to $8$, as one can observe at least eight different couples. For each couple~$n$, one has
\begin{equation}
\Omega_{1,n}+\Omega_{2,n}=\Omega_0.
\end{equation}
The index $n$ classifies the different couples from the most intense ($n=1$) to the less intense ($n=8$).

The first couple is generated by the attractor itself, as a standard TRI described in~\cite{SED2013} or in \S~\ref{TRIpropagating}. Then, other couples appear and all the frequencies in the tank are linked by $3$-wave interactions. The cascade is created as follows. Once the first couple has been created by the attractor, the two frequencies of this couple \ADD{($\Omega_1=\Omega_{1,1}$ and $\Omega_2=\Omega_{2,1}$)} interact together to create a third frequency \ADD{($\Omega_{2,2}$)}, which belongs to the second couple. The second couple is completed by the interaction between the third frequency and the frequency of the attractor, $\Omega_0$. These reactions are given by
\begin{eqnarray}
\Omega_{1,1}-\Omega_{2,1}&=&\Omega_{2,2},\\
\Omega_{0}-\Omega_{2,2}&=&\Omega_{1,2}.
\end{eqnarray}
Thus, the second couple is complete. Now, several interactions are possible between the different frequencies. An interaction between the first and the second couples leads to a third one:
\begin{eqnarray}
\Omega_{1,2}-\Omega_{1,1}&=&\Omega_{2,3},\\
\Omega_{0}-\Omega_{2,3}&=&\Omega_{1,3},
\end{eqnarray}
and an interaction between the frequencies of the second couple creates a forth couple:
\begin{eqnarray}
\Omega_{1,2}-\Omega_{2,2}&=&\Omega_{1,4},\\
\Omega_{0}-\Omega_{1,4}&=&\Omega_{2,4}.
\end{eqnarray}
Here, one has four different couples, so eight frequencies. One can easily continue to create other frequencies by combining the different frequencies of the different couples. This mechanism leads to an infinite set of discrete frequencies and the first eight couples are at least visible in figure~\ref{20140513_tps_freq}. \ADD{Note that each coupling satisfies also the spatial resonant condition and that all the waves fulfill the dispersion relation.} At the end, each frequency is linked with a large number of other frequencies by a three-wave interaction~(\cite{BrouzetEPL2016}). Experimentally, we observe the fullfilment of the above equations with the accuracy in non-dimensional frequency of $\pm 0.001$.

\subsection{Observational evidence of standing secondary  waves}

Although the instability starts and evolves differently in well- and weakly-focused attractors, they can exhibit some common features. Figure~\ref{20140619_filles} shows that the secondary waves generated by the weakly-focused attractor are standing waves. Some frequencies created by a well-focused attractor in long-term experiments can also be standing waves as shown in figure~\ref{20140513_filles}. For example, the fluid motions at frequencies $\Omega_{2,1}$ and $\Omega_{2,2}$ are standing waves as 
it is clear from the wave patterns filtered at these frequencies. Note that not all the frequencies present in figure~\ref{20140513_tps_freq} correspond to standing waves.

\begin{figure}
   \begin{centering}
 \includegraphics[width=0.8\textwidth]{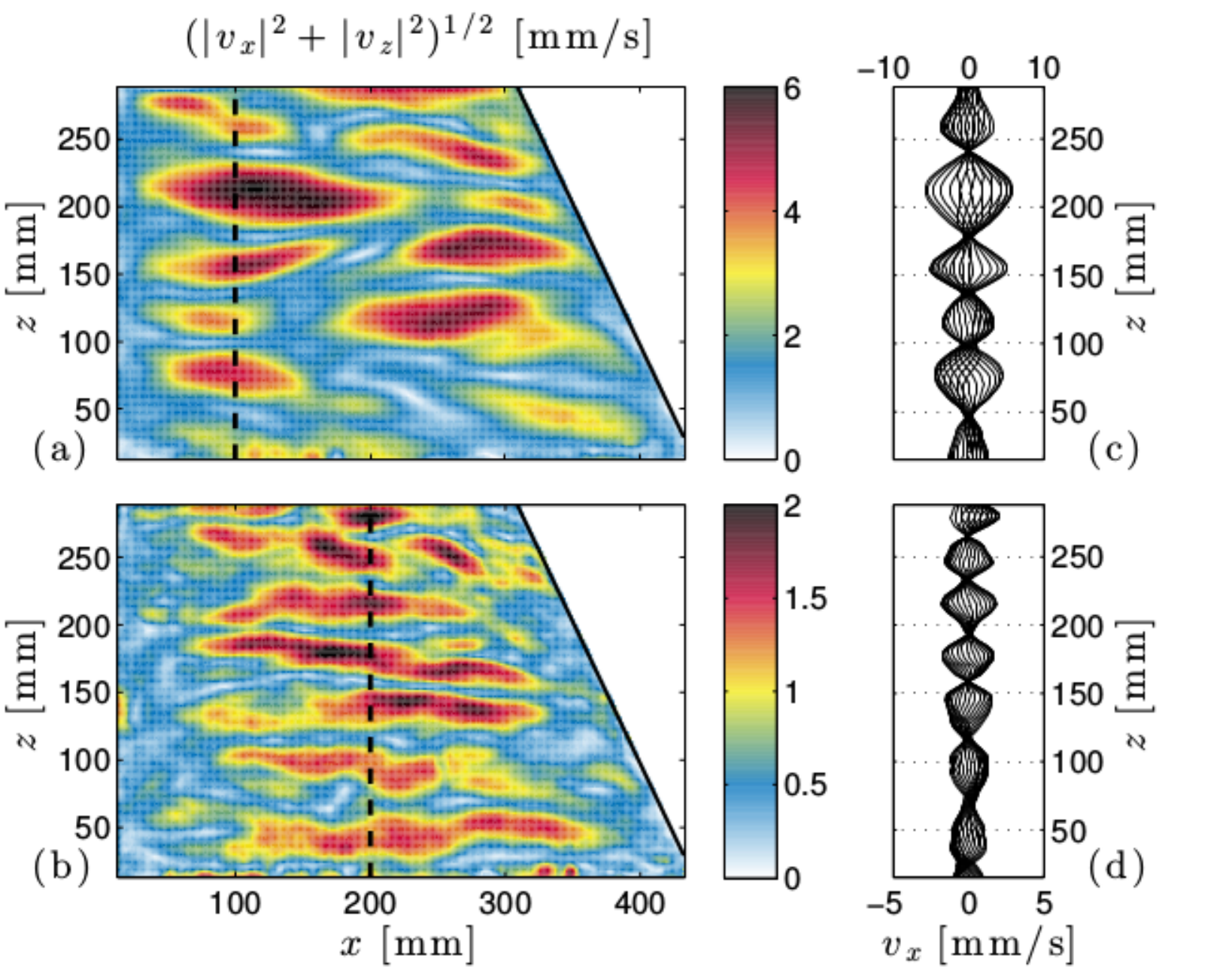}
   \caption{Components of the wave field corresponding to frequencies $\Omega_{2,1}$ (a) and $\Omega_{2,2}$ (b), obtained with the Hilbert transform around $620~T_0$ using a time window $85~T_0$ long. Note that the components oscillating at $\Omega_{2,1}$ and $\Omega_{2,2}$  correspond to standing waves as is clearly seen on the sequences of wave profiles shown on the right of each picture. Results are presented for the experiment indicated by the symbol $\blacksquare$ in figure~\ref{diagram} {with $a=5$~mm}
    (see Table~\ref{tabular:parametersMainexperiments} for the parameters).     \label{20140513_filles}}
\end{centering}
\end{figure}

With the frequencies corresponding to the peaks of the spectrum shown in figure~\ref{20140513_tps_freq}(b), one can compute the values of parameters $\tau$ associated to these waves. These frequencies are shown by different symbols in the diagram $(d,\tau)$ presented in figure~\ref{nouveaudiagram}. 
It can be seen that the $\tau$ parameters for the standing secondary waves are close to black zones in the $(d,\tau)$ diagram.   Thus, at least some frequencies created by the weakly- and well-focused attractors could correspond to quasi-resonant modes.  In absence of commonly accepted terminology, we can tentatively call the dark regions (see figure~\ref{nouveaudiagram}) as geometric quasi-resonances.
{Previously quasi-resonances have been discussed  assuming non-zero resonance width (see~\cite{Nazarenko2011}). Here we refer to possibility of global quasi-resonances assuming that the Lyapunov exponent can take very small negative non-zero value.}
 In a realistic system with weak viscous dissipation, we can expect that some of the quasi-resonances can exhibit a seiche-like behavior
{similar to exact seiche modes discussed in~\cite{MBSL1997} and~\cite{MandersMaas2003}}. Therefore, at the conceptual level, the key driving processes in long-term experiments with unstable wave attractors can be considered as a combination of "local" \citep{SED2013} and "global-scale" (present paper) TRI events, with long-term transients between the two.

Interesting parallels can be found in the dynamics of rotating fluids.  For instance, \ADD{\cite{Beardsley1970}, \cite{Thompson1970}} and \cite{DuguetScottLePenven2006} show rich multi-peak spectra for inertial oscillations in a compressible fluid confined in a rotating cylindrical vessel; they relate the frequencies associated with the observed peaks to global modes of the fluid motion. Similarly, \cite{Favier-etal-2015} analyse the frequency spectrum of fluid motion generated by libration-driven elliptic instability in a rotating ellipsoid and show the presence of the eigenfrequencies of linear and quadratic inertial modes.

\section{The ``wave mixing box" regime in NP and NS attractors} 

In the previous Sections, we have shown that the \ADD{scenario} of instability in wave attractors and the parameters of wave triplets involved into resonant interactions strongly depend on the choice of the operating points in $(d, \tau)$ plane. However, one can expect a certain level of universality \ADD{for fully developed cascades of triadic interactions transferring energy to small-scale internal waves and ultimately to mixing events. A high mixing efficiency of triadic cascades operating in confined domains has been demonstrated so far in~\cite{McEwan1983}  for normal modes and in ~\cite{BrouzetEPL2016} for well-focused attractors, in rectangular and trapezoidal domains, respectively.}
 
Below,  we compare the case of well-focused $(1,1)$ attractor described in~\cite{BrouzetEPL2016} 
($\blacksquare$) with the case of a weakly focused $(1,1)$ attractor ($\bigstar$). The time-frequency diagrams of two unstable attractors \ADD{obtained at high amplitude of forcing ($a=10$~mm) in regimes with measurable mixing}  are shown in figure \ref{time-frequency-mix}.  Both diagrams have qualitatively similar features: several discrete frequency peaks of large magnitude are embedded into a continuous \ADD{frequency} spectrum, which has significantly smaller magnitude. The \ADD{magnitudes of discrete} peaks \ADD{strongly} fluctuate in time. 

\begin{figure}
\begin{centering}
\includegraphics[width=0.9\textwidth]{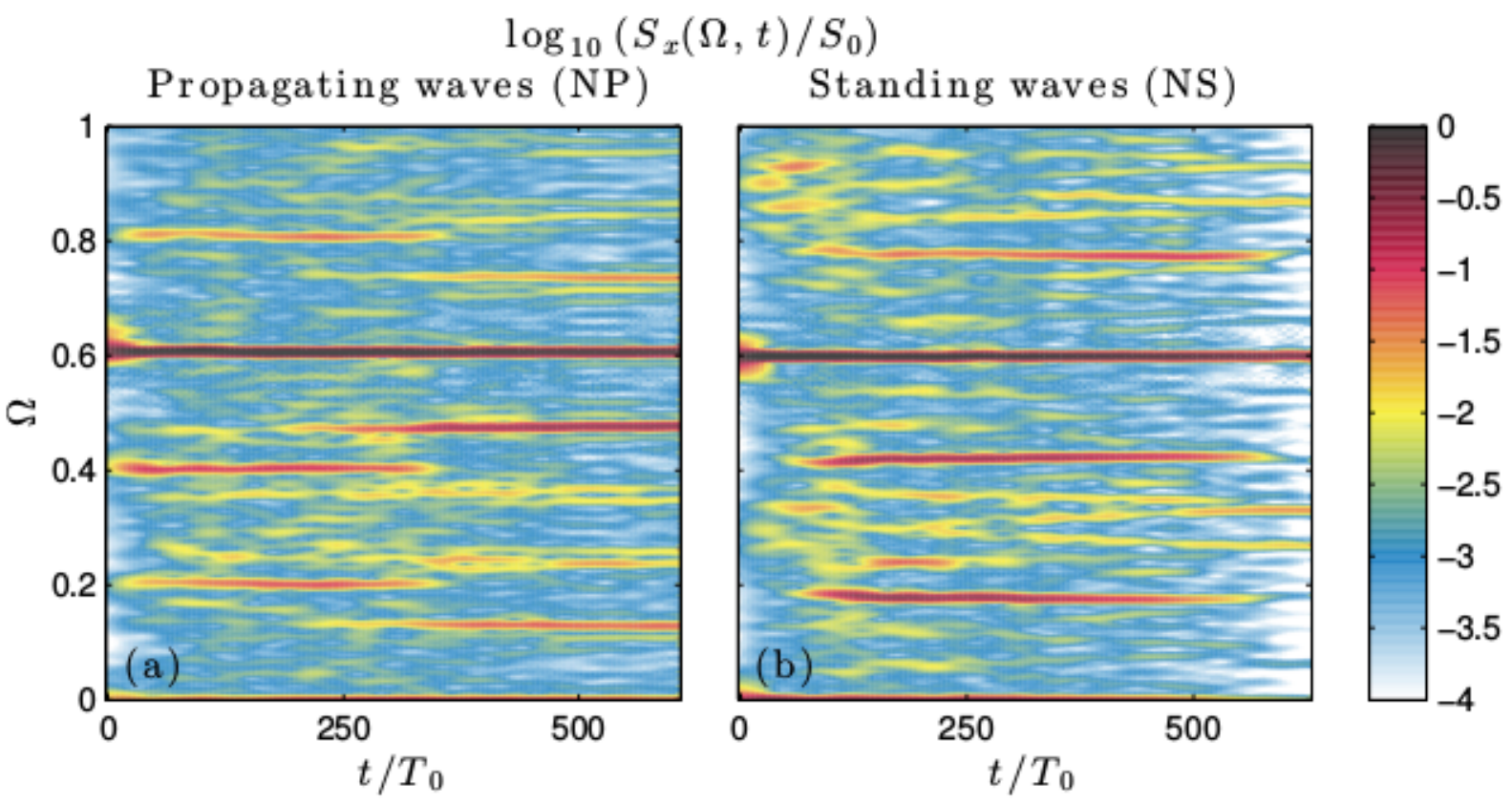}
\caption{Time-frequency diagrams of two signals recorded in unstable wave attractors corresponding to different operating points. Both attractors are unstable and the wave regime is such that it induces significant mixing.
The left panel corresponds to the experiment indicated by the symbol $\blacksquare$ in figure~\ref{diagram},
while the right one by the $\bigstar$ 
(with $a=10$~mm and other parameters listed in
Table~\ref{tabular:parametersMainexperiments}).
\label{time-frequency-mix}}
\end{centering}
\end{figure}

\cite{BrouzetEPL2016} show that the normalized horizontal $y$-component of the vorticity field, 
\begin{equation}
\Xi_{y}=\frac{\xi_{y} (x,z,t)}{N}=\frac{1}{N}\left( \frac{\partial v_x}{\partial z}-\frac{\partial v_z}{\partial x} \right) ,
\end{equation} 
represents a \ADD{useful} quantity \ADD{indicating the occurence}  of extreme events in the experimental system. 
This quantity has a clear physical meaning as a ratio of the destabilizing effect of vorticity to the stabilizing effect of stratification. Obviously, we can expect some mixing in the system when a significant statistics of events with high values of~$|\Xi_{y}|$ is \ADD{observed}. 

Figure~\ref{PDF_vorticity} represents the \ADD{probability density functions (PDFs)} of $\Xi_{y}$ for $(1,1)$ attractors with weak and strong focusing for two different values of the forcing amplitude $a$. 
 \ADD{Comparing the data obtained at fixed $a$, we see that quantitatively} the system with weak focusing  shows consistently lower statistics of extreme events \ADD{than the system with high focusing.} \ADD{Qualitatively, the shape of vorticity PDFs in attractors with high and low focusing is} similar.   Moreover, the long-term \ADD{erosion of} stratification is also similar \ADD{as discussed below}. 

The significant statistics of extreme events with high horizontal vorticity arises from the spontaneous summation of the frequency components of the wave field. \ADD{It is important to note} that the primary \ADD{(monochromatic)} wave alone cannot produce \ADD{extreme} events.  This is attested by figures~\ref{PDF_vorticity}\ADD{(c) and (d)} that represent the PDFs of $\Xi_{y}$ for the primary wave field filtered at~$\Omega_{0}$. It can be seen that these PDFs have no tails extending to the domains with $|\Xi_{y}|>2$. The threshold value $|\Xi_{y}|>2$ has been  \ADD{proposed} in \cite{BrouzetEPL2016} as extension of the classic Miles-Howard criterion for stability of horizontal shear flows of continuously stratified fluid. In fact, \ADD{a possibility of extension of the Miles-Howard criterion to a wider context}  \ADD{has been discussed in some} classic works, e.g. \cite{Phillips}. \ADD{Using}  $|\Xi_{y}|=2$ \ADD{as a threshold value}  \ADD{one can} make some qualitative conclusions. 

In figure\ADD{s}~\ref{PDF_vorticity}(e) and~(f), we show the \ADD{ratios of the final to initial stratifications} observed in different experiments described in the present paper. 
\ADD{In both cases at large amplitude of forcing} we observe a well-measurable mixing when comparing the initial and final density distributions. 
The probability of having an extreme event with intensity $|\Xi_{y}|>2$ is given by the integral over the corresponding tails of the PDF. The cases with measurable mixing depicted in figure\ADD{s}~\ref{PDF_vorticity}(e) and~(f) correspond to drastically increased probability of extreme events. It increases
from 2\textperthousand\, for $a=$ 5~mm to 19\textperthousand\, for $a=$ 10 mm in the "propagating waves" case,
and  from 0.7\textperthousand\, for $a=$~5 mm to 5.6\textperthousand\, for $a=$ 10 mm in the "standing waves" case.
We observe a comparable increase of probability, by factor $10$ and $8$, in both cases. {Although the statistics of high-vorticity events seems to represent a kinematic indicator for the occurence of mixing, at the present stage it is difficult to relate this statistics directly to the overall mixing efficiency. Additional parameters should be measured, most importantly, the typical scale of overturning structures. As an alternative, one can make an attempt to estimate the turbulent diffusion coefficient in the experimental system similar to~\cite{PIVLIF}.}

\begin{figure}
   \begin{centering}
  \includegraphics[width=0.75\textwidth]{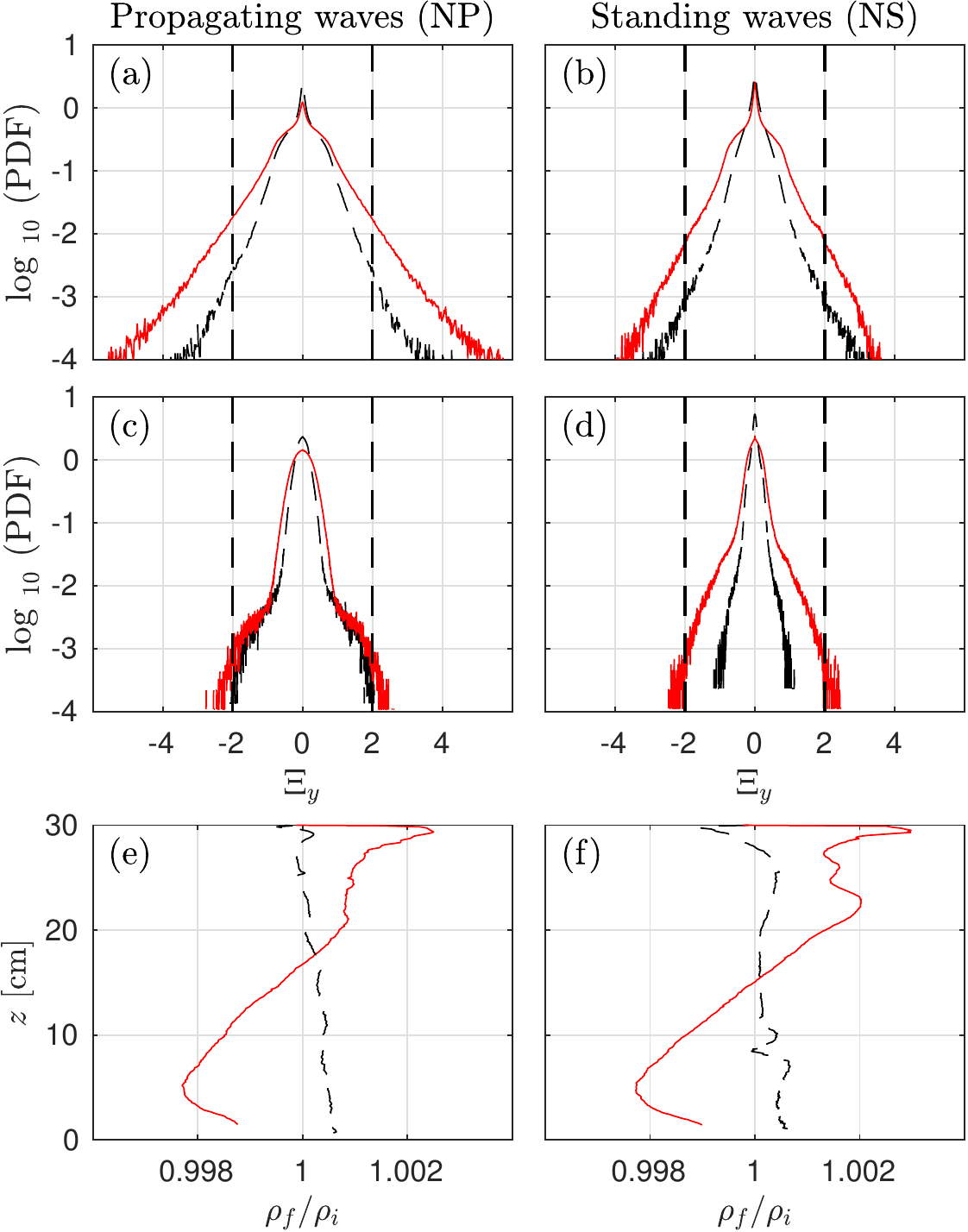}
   \caption{Panels (a) to (d) present the probability density functions of the vorticity in the tank, while
  panels~(e) and (f) exhibit the ratio of the final (after 700 periods) to initial stratifications.  
  The PDF are calculated from all images between $480$ and $500~T_0$, for the two propagating (panel a, {$\blacksquare$ in figure~\ref{diagram}}) or standing (panel b, {$\bigstar$ in figure~\ref{diagram}}) wave experiments presented in figure~\ref{time-frequency-mix}. Panel (c) and~(d)
present the same quantity for the vorticity 
  filtered at $\Omega_0$. {\ADD{In all panels,} dashed black and solid red lines correspond to $a=5$~mm and $a=10$~mm.}
  \label{PDF_vorticity}}
\end{centering}
\end{figure}

\section{Conclusions}

Geometric focusing of internal wave motions in a confined fluid domain leads to energy concentration at a closed loop, an internal wave attractor (Maas \& Lam 1995, Maas et al. 1997). The rate of convergence of internal wave rays toward limiting cycle can be quantified in terms of Lyapunov exponents. Using this variable, one can define the domains of existence (Arnold tongues) of wave attractors. 

In the present paper, we explore the structure of the Arnold tongue corresponding to a simple parallelogram-shaped $(1,1)$ attractor in a trapezoidal domain of a uniformly stratified fluid. We show that the kinetic energy of the confined fluid system represents an appropriate global variable that allows to classify the observed wave regimes. A typical Arnold tongue is shown to have a central region corresponding to "classic" {well focused} wave attractors with thin branches where the wave energy is concentrated. These attractors typically have a large total kinetic energy, and the wave motion in the attractor branches is represented by a \ADD{propagating} wave. The behaviour of "classic" wave attractors has been studied in great details, both in linear~\citep{MaasLam1995,MBSL1997,RGV2001,Ogilvie2005,HBDM2008,GSP2008,HTMD2008,EYBP2011,GuoHolmesCerfon2016} and non-linear~\citep{Ogilvie2005,SED2013,JouveOgilvie2014} regimes including the onset of triadic resonance and formation of energy cascades of triadic wave-wave interactions.

Importantly, a typical Arnold tongue has border regions corresponding to zones of geometric degeneracy which can be of two types: i) attractor collapses onto a line {representing a diagonal of the trapezoid} \ADD{(see figure~\ref{nouveaudiagram})} or ii) the system itself does not support effective focusing. \ADD{In this latter case, each internal wave orbit is periodic contrary to attracting cases, for which only a few periodic orbits exist and are named attractors. Moreover, this} case corresponds, as a limit, to well-studied configuration of standing internal waves in a rectangular domain filled with uniformly stratified fluid~\citep{McEwan1971}. 
The intermediate case of an elongated attractor in the domain with weak focusing is studied in detail in the present paper. For this case, the wave motion in the attractor 
 \ADD{has both the properties of standing and propagating waves. Importantly, }increased energy concentration at the loop predicted by the ray tracing indicates that the \ADD{propagating}  component of the wave field does matter. 
We show that in this configuration under appropriate forcing one can observe a scenario of triadic resonance with very low growth rate of secondary waves: the exponential growth for nearly three decades in amplitude at the time-scale of one hour has been detected experimentally. The secondary waves are found to be represented by \ADD{nearly} standing waves. Thus, at the border regions of the Arnold tongue the onset of triadic instability represents an intermediate case between the cases of purely \ADD{propagating}~\citep{SED2013} and purely standing~\citep{McEwan1971} wave triplets.

Further, we explore the long-term behaviour of instability detected in~\cite{SED2013} and \cite{BrouzetEPL2016}. We show that some frequencies in the spectrum of wave motions correspond to standing and nearly-standing waves with high vertical modes. These observations may have an important consequence to the dynamics of large stratified geophysical systems. Such geophysical systems (lakes for instance) may be subject to very complex forcing. The cascade of triadic instabilities that develops as a response to forcing may effectively transfer energy to standing and nearly-standing waves of high vertical modes. When the forcing stops, the attractor-like components of the internal wave motion quickly disappear as described in \cite{HBDM2008} and \cite{GSP2008}. In absence of energy injection, at low wave numbers, the spectrum of wave motions in attractors quickly shifts toward high wave numbers due to focusing
, and the high wave numbers quickly decay due to viscosity. In contrast, the normal vertical (global resonance) modes conserve their length scale and decay at much longer time-scale owing to purely viscous mechanism. Similar dynamics is expected for quasi-resonant modes characterized by vanishingly weak focusing
. Such a scenario can explain the presence of high vertical modes in limnological observations.

As the energy input into a confined domain increases, one can expect a cascade of energy from global to small scales. As it is shown for well-focused "classic" wave attractors~\citep{BrouzetEPL2016}, such a cascade operates via \ADD{a} hierarchy of triadic wave-wave interactions transferring energy to small scale dissipation and mixing events. We show that under sufficiently large forcing this is a common fate for wave attractors with strong and weak focusing, with qualitatively similar statistics of extreme events leading to mixing. Interestingly, even in the limit of rectangular box the cascade of triadic interactions can transfer a significant portion of energy to mixing as shown in~\cite{McEwan1983}. The choice of the operating point \ADD{inside} the Arnold tongue and the amplitude of the input perturbation provide extreme flexibility to control the parameters of the energy cascade in the system. We believe that this "mixing box" configuration has a strong potential to mimic many aspects of the energy cascade in confined geophysical systems both experimentally and numerically.  

\ADD{Interestingly, in the end, 
both well and weakly focused attractors 
display similar mixing properties. 
This generic result is of importance to argue that any continuously-stratified and continuously-forced fluid 
repeatedly goes through a sequence of focusing, mixing and restratification. 
The presence of non-uniform stratification and mean flows
has sometimes been interpreted as presenting counter-evidence of the occurrence 
of wave attractors~\citep{GerkemavanHaren2012}. 
 In the light of the present result, 
it may be worth to reconsider this interpretation.}

In the present paper, we have \ADD{measured} the \ADD{kinetic energy of stable attractors} \ADD{as function of their location at} the Arnold tongue at the laboratory scale. At large scale we expect the increase of the central region of the Arnold tongue corresponding to classic "narrow-beam" attractors with \ADD{propagating} wave motion, and decrease of the border regions corresponding to attractors with considerable standing-wave component. However, such a {qualitative} extrapolation to large scales applies only to stable attractors in linear viscous regime \citep{HBDM2008,GSP2008}.
Recent results presented in~\cite{GuoHolmesCerfon2016} suggest that for typical bathymetry of the ocean bottom in a two-dimensional model, one can expect about 10 attractors per 1000 km. Interestingly, this study shows that in the case of small-scale bathymetry the attractors with very elongated loops represent the most probable configuration, justifying our interest to attractors with a “degenerated” geometry.
The realistic wave regime in large-scale stratified reservoirs {in presence of wave attractors} is likely to be nonlinear, with signatures of internal-wave turbulence and mixing as described in \cite{BrouzetEPL2016} and the present paper. {The relevance of such a scenario to a three-dimensional topography \cite{MandersMaas2004,DM2007,WZLDQ2015} and to inertial-wave attractors in spherical liquid shells \cite{Favier-etal-2014} remains to be explored.}

\begin{acknowledgments}
{\bf Acknowledgments}

EVE gratefully acknowledges his appointment as a Marie Curie incoming fellow
at Laboratoire de physique ENS de Lyon. 
This work was supported by the LABEX iMUST (ANR-10-LABX-0064) of Université de Lyon, within 
the program "Investissements d'Avenir" (ANR-11-IDEX-0007) operated by the French National 
Research Agency (ANR). This work has 
achieved thanks to the resources of PSMN from ENS de Lyon.
We thank L. Maas and H. Scolan for helpful discussions.

\end{acknowledgments}

\end{document}